\documentclass[draft]{agujournal2019}
\usepackage{url} %this package should fix any errors with URLs in %refs.
\usepackage{lineno}
\usepackage[inline]{trackchanges} %for better track changes. finalnew option will compile document with changes incorporated.
\usepackage{soul}
\usepackage{amsmath}
\usepackage{lmodern}
\usepackage{float}
\draftfalse

\journalname{Journal of Advances in Modeling Earth Systems (JAMES)}

\begin{document}

\title{A simple emulator that enables interpretation of parameter-output relationships, applied to two climate model PPEs}

%%%%%%%%%%%%%%%%%%%%%%%%%%%%%%%%%%%%%%%%%%%%%%%
%
%  AUTHORS AND AFFILIATIONS
%
%%%%%%%%%%%%%%%%%%%%%%%%%%%%%%%%%%%%%%%%%%%%%%%

% Authors are individuals who have significantly contributed to the
% research and preparation of the article. Group authors are allowed, if
% each author in the group is separately identified in an appendix.)

% List authors by first name or initial followed by last name and
% separated by commas. Use \affil{} to number affiliations, and
% \thanks{} for author notes.
% Additional author notes should be indicated with \thanks{} (for
% example, for current addresses).

\authors{Qingyuan Yang\affil{1,2}, 
         Gregory S Elsaesser\affil{1,3,4},
         Marcus Van Lier-Walqui\affil{1,3,5},
         Trude Eidhammer\affil{6}}

\affiliation{1}{Learning the Earth with Artificial Intelligence and Physics (LEAP) National Science Foundation (NSF) Science and Technology Center, Columbia University, New York, NY, USA}
\affiliation{2}{Department of Earth and Environmental Engineering, Columbia University, New York, NY, USA}
\affiliation{3}{NASA Goddard Institute for Space Studies, NY, USA}
\affiliation{4}{Department of Applied Physics and Applied Mathematics, Columbia University, New York, NY, USA}
\affiliation{5}{Center for Climate System Research, Columbia University, New York, NY, USA}
\affiliation{6}{NSF National Center for Atmospheric Research, Boulder, CO, USA}

%\affiliation{=number=}{=Affiliation Address=}
%(repeat as many times as is necessary)

% Corresponding author mailing address and e-mail address:

% (include name and email addresses of the corresponding author.  More
% than one corresponding author is allowed in this LaTeX file and for
% publication; but only one corresponding author is allowed in our
% editorial system.)

% Example: \correspondingauthor{First and Last Name}{email@address.edu}

\correspondingauthor{Qingyuan Yang}{qy2288@columbia.edu}

%%%%%%%%%%%%%%%%%%%%%%%%%%%%%%%%%%%%%%%%%%%%%%%
% KEY POINTS
%%%%%%%%%%%%%%%%%%%%%%%%%%%%%%%%%%%%%%%%%%%%%%%
%  List up to three key points (at least one is required)
%  Key Points summarize the main points and conclusions of the article
%  Each must be 140 characters or fewer with no special characters or punctuation and must be complete sentences

% Example:
% \begin{keypoints}
% \item	List up to three key points (at least one is required)
% \item	Key Points summarize the main points and conclusions of the article
% \item	Each must be 140 characters or fewer with no special characters or punctuation and must be complete sentences
% \end{keypoints}

\begin{keypoints}
\item A simplified Additive Gaussian Processes method is designed for emulating sparsely-sampled climate model perturbed parameter ensembles (PPEs).
\item The method, tested on NCAR CAM and GISS ModelE PPEs, exhibits performance similar to other emulators, but with less dependence on hyperparameters.
\item The method provides new diagnostics on parameter–output importance not captured in linear correlation maps, enabling new insights that could benefit the future design of PPEs and emulators.
\end{keypoints}

%%%%%%%%%%%%%%%%%%%%%%%%%%%%%%%%%%%%%%%%%%%%%%%
%
%  ABSTRACT and PLAIN LANGUAGE SUMMARY
%
% A good Abstract will begin with a short description of the problem
% being addressed, briefly describe the new data or analyses, then
% briefly states the main conclusion(s) and how they are supported and
% uncertainties.

% The Plain Language Summary should be written for a broad audience,
% including journalists and the science-interested public, that will not have 
% a background in your field.
%
% A Plain Language Summary is required in GRL, JGR: Planets, JGR: Biogeosciences,
% JGR: Oceans, G-Cubed, Reviews of Geophysics, and JAMES.
% see http://sharingscience.agu.org/creating-plain-language-summary/)
%
%%%%%%%%%%%%%%%%%%%%%%%%%%%%%%%%%%%%%%%%%%%%%%%

%% \begin{abstract} starts the second page

\begin{abstract}
We present a new additive method, referred to as \textit{sage} for Simplified Additive Gaussian processes Emulator, for emulating climate model Perturbed Parameter Ensembles (PPEs). \textit{sage} estimates the value of a climate model output as the sum of additive terms. Each additive term is the mean of a Gaussian Process, and corresponds to the impact of a parameter or parameter group on the variable of interest. This design caters to the sparsity of PPEs which are characterized by limited ensemble members and high dimensionality of the parameter space. \textit{sage} quantifies the variability explained by different parameters and parameter groups, providing additional insights on the parameter-climate model output relationship. We apply \textit{sage} to two climate model PPEs and compare it to a fully connected Neural Network. The two methods have comparable performance with both PPEs, but \textit{sage} provides insights on parameter and parameter group importance as well as diagnostics useful for optimizing PPE design. Insights gained are valid regardless of the emulator method used, and have not been previously addressed. Our work highlights that analyzing the PPE used to train an emulator is different from analyzing data generated from an emulator trained on the PPE, as the former provides more insights on the data structure in the PPE which could help inform the emulator design. 
\end{abstract}

\section*{Plain Language Summary}
Climate models have many parameters whose values are uncertain. A parameter could be one that affects the fall speed of snow in the climate models. Researchers design Perturbed Parameter Ensembles (PPEs) to study how the parameters affect the model output (e.g., precipitation) so as to gain insight on the best settings for parameters. A PPE is a set of climate model runs (i.e., the ensemble members; normally a few hundred) with different parameter values. Since running climate models is very computationally expensive, all parameter combinations cannot be run.  In this work, we propose a "surrogate model" or "emulator" method that, once properly trained, estimates the climate model output climatologies given a set of model parameters that have not been run by the climate model. The method is additive, meaning that the impacts of parameters and parameter interactions on the climatologies are assumed additive. The method is unique in that it analyzes the relationship between the parameters and climatologies in a PPE and makes predictions at the same time. We apply our method to two climate model PPEs, and compare its performance with that of fully connected Neural Networks. The performance of the two methods is comparable, and additional insights can be obtained from using our methods. Such insights are universal, and not method-specific. Our work highlights that analyzing the PPE data is different from analyzing data generated from an emulator trained on the PPE.

%%%%%%%%%%%%%%%%%%%%%%%%%%%%%%%%%%%%%%%%%%%%%%%
%
%  BODY TEXT
%
%%%%%%%%%%%%%%%%%%%%%%%%%%%%%%%%%%%%%%%%%%%%%%%

%%% Suggested section heads:

\section{Introduction}
Earth System Model (ESM) perturbed parameter ensembles (PPEs) are useful tools for analyses of ESM emergent property or performance score sensitivities to physics parameter choices (e.g., \citeA{murphy2004quantification,collins2011climate,williamson2013history,parker2013ensemble,haughton2014generation, sexton2019finding, duffy2024perturbing,eidhammer2024extensible, elsaesser2024,gettelman2024interaction}) as well for broadly quantifying envelopes of projection uncertainty \cite{hourdin2023toward}. PPEs are generated by running a climate model with different sets of parameter values. Latin Hypercube Sampling (LHS; \citeA{mckay2000comparison}), which is similar to random sampling but ensures  that the sampled parameter marginal distributions are uniform, is commonly used for generating the parameter values (e.g., \citeA{lambert2013interactions, li2019reducing,eidhammer2024extensible}), although there are other more sophisticated sampling approaches \cite{dunbar2021calibration, cleary2021calibrate,elsaesser2024}.

PPEs commonly serve as the training data for building climate model emulators (e.g., \citeA{hawkins2019parametric,sexton2019finding,dagon2020machine,
tran2021development,gettelman2022future,beucler2022machine, elsaesser2024}). Once properly trained, an emulator can approximate the outputs of a computationally expensive climate model at a negligible cost. The emulator can then propagate climate model outputs for various purposes such as parameter estimation \cite{hourdin2017art,li2019reducing,dagon2020machine,van2020bayesian, watson2021model,fletcher2022toward,peatier2022investigating,carzon2023statistical,elsaesser2024}, sensitivity analysis \cite{webb2013origins,sexton2019finding,proske2022assessing}, and uncertainty propagation \cite{hutchins2016projections,hourdin2023toward}.

Different methods have been developed for analyzing and emulating climate model PPEs. These include linear regression (e.g., \citeA{duffy2024perturbing}), Gaussian Processes (GPs; \citeA{lee2011emulation,regayre2014uncertainty,johnson2015evaluating,yang2020novel,carzon2023statistical}) and Neural Network (NN; \citeA{reichstein2019deep,labe2021detecting,dagon2022machine}) frameworks. Different emulation methods have also been integrated into a single framework, providing the flexibility of choosing the optimal method after inter-comparison (e.g., \citeA{watson2021model}) as well as providing information on emulator structural uncertainty.

With the increasing use of PPEs, we need not only reliable emulators but also tools that analyze some of the nuanced (and non-linear) structure existing in climate model PPEs. This is an important consideration as more sophisticated analysis tools not only aid in PPE interpretation and emulation, they also can inform on future design and sampling considerations for subsequent PPEs, particularly given the known limitations with sampling in high dimensional state spaces. Additionally, traditional approaches for examining the parameter-target variable (i.e., the model output or model skill score) relationships have limitations. For example, the popular ``heat map'' revealing the pairwise correlation between parameters and target variables does not consider the effect of parameter interaction. 
In choosing and designing an emulator, we cannot easily and explicitly explain how emulator performance is affected by different qualities (e.g., the number of ensemble members) of the analyzed PPE.  For example, we do not know whether emulating the target variables all at once or a group at a time impacts emulator performance without brute force manual testing. A tool that reveals PPE structure could help address the above questions. It is worth noting that analyzing a PPE is different from analyzing a large number of samples generated from an emulator that is trained based on the PPE, because the generated samples are subject to additional errors from the imperfect emulation, and since this additional error is often not considered in subsequent analyses, investigators are likely to over-interpret the corresponding results.

Toward addressing the unknowns listed above, in this work we propose a simple method that emulates and provides useful analysis of climate model PPEs simultaneously. We compare the performance against a NN that was recently used in studies pertaining to two diverse PPEs \cite{eidhammer2024extensible, elsaesser2024}. Our new method captures the general structure (i.e., the non-local and low-frequency relationship in the state space) between parameters and target variables (analogous to a low-pass filter). It is designed to cater to the sparsity (i.e., small ensemble size relative to the number of parameters) of typical climate model PPEs.

The method is simplified from additive Gaussian Processes \cite{plate1999accuracy, duvenaud2011additive,cheng2019additive} but with modifications that enable improved performance for sparsely sampled state spaces. We term the method \textit{sage} for Simplified Additive Gaussian processes Emulator. We refer to the method \textit{sage} or ``the (present or current) method'' interchangeably in the following text. The method resembles the polynomial-based emulator described in \citeA{williamson2013history} in its way of selecting the effective parameters. It sequentially and iteratively identifies the parameters and parameter groups that a target variable is sensitive to, and models their impact on the target variable. The sum of these impacts (i.e., additive terms) represents the emulator prediction.

\textit{sage} estimates the target variable as the sum of multiple terms, each corresponding to a parameter or parameter groups. It quantifies and outputs how much the Root-Mean-Square Error (RMSE) is gradually reduced after each term is added to the emulator prediction for training and validation datasets. The reduced RMSE quantifies how individual parameters and parameter groups contribute to the emulator performance, providing insights on their impact on the target variable.

We apply \textit{sage} to two PPEs generated from different climate models: (1) the latest Goddard Institute for Space Studies (GISS) ESM ModelE3 \cite{cesana2019evaluating,russotto2022evolution,elsaesser2024} and (2) the Community Earth System Model version 2 (CESM2; \citeA{danabasoglu2020community}) Community Atmosphere Model 6 (CAM6; \citeA{gettelman2019high}). We compare emulator performance against a NN emulator used in studies of the same two PPEs (i.e., from \citeA{elsaesser2024} and \citeA{eidhammer2024extensible}), before extension to the PPE-structural-analysis benefit provided by \textit{sage}. Our work reveals previously-unnoticed insights that would benefit future PPE analysis and emulator design. Highlights of such insights include: (1) neither emulating the variables all at once nor one-at-a-time is (always) the perfect emulator design; (2) a group of less-sensitive parameters could have a non-negligible and cumulative impact on the overall emulator performance; (3) we need to be \textit{very} careful in determining whether a parameter is important or not; and (4) the emulator performance does not not change linearly with the number of ensemble members for training.

The analysis, comparison, and findings based on the application of two emulator methods (\textit{sage} and a NN) to two different PPEs are considered equally or even more important than the presented method itself, as they inform the design of PPEs and climate model emulators for future studies regardless of the emulator method used.

\section{Datasets}
We leverage two very different PPEs in this work to demonstrate the value of \textit{sage}.  The two PPEs are fully described in \citeA{elsaesser2024} and \citeA{eidhammer2024extensible}, respectively. They are briefly introduced here.

The ModelE3 PPE has 851 members, with each member characterized by a unique set of 45 physics parameters spanning the atmospheric convection, large-scale cloud, and turbulence modules of ModelE3. Their ranges, summarized in Table 1 of \citeA{elsaesser2024}, were determined based on modeler expertise as in \citeA{sexton2021perturbed}. Each parameter combination is used in a short 1-year, atmosphere-only simulation (110 vertical layers, 2.5 $\times$ 2.0$^\circ$ horizontal resolution) run with a present-day prescribed sea surface temperature climatology and atmospheric composition. The first 451 (450 plus one default run) PPE members were generated using parameter combinations provided by LHS. The parameters of the next 100 PPE members were generated using random sampling. The final 300 sets of parameters were sampled \textit{sequentially} (100 at a time for three times) based on a method detailed in \citeA{elsaesser2024} that aims to generate a special type of PPE (termed a ``calibrated physics ensemble'', or CPE) whose members agree with numerous earth energy and water cycle diagnostics. This 300 member CPE relied on machine learning and a Markov Chain Monte Carlo (MCMC) method to sample from a target probability distribution defined by comparison of the emulated model with information from numerous observational products. As a result, relative to the first 551 members, the three 100-member batches yield outputs exhibiting progressively reduced biases with respect to satellite observations. In this work, we exclude the last 100 ensemble members in the event that they are too clustered near an hypervolume in the parameter space, which might affect the emulator performance (i.e., the output variables of these 100 ensemble members tend be very similar with each other, hence whether and how many of them are included in the training and validation datasets would greatly affect the method performance).

CESM2 \cite{danabasoglu2020community} CAM6 \cite{gettelman2019high} serves as the base model for generation of the second PPE used in this work. Hereafter, we often refer to this PPE as the CAM6 PPE. The CAM6 PPE has 262 members with 45 parameters perturbed and sampled using LHS. These parameters span the unified turbulence closure (CLUBB; \citeA{golaz2002pdf}), cloud microphysics (MG2; \citeA{gettelman2015advanced}), Modal Aerosol Model (MAM; \citeA{liu2012toward}), and the Zhang-McFarlane deep convection (ZM; \citeA{zhang1995sensitivity}) schemes. Their ranges are determined with expert elicitation, and can be found in Table 1 of \cite{eidhammer2024extensible} (also not listed here for simplicity). Two pairs ($clubb\_C6rt$ and $clubb\_C6thl$; $clubb\_C6rtb$ and $clubb\_C6htlb$) of CLUBB parameters are perturbed and sampled simultaneously, reducing the number of free parameters to 43. Each PPE member simulation is run using near present day cyclic boundary conditions for year 2000. The greenhouse gases and atmospheric oxidants take the average values of the 1995-2005 period. The average monthly sea surface temperatures (SSTs) during 1995-2010 are used. The emission of aerosols and precursors is set to 1995-2005 in these present day simulations. The simulation resolution is $0.9^\circ$ (latitude) $\times 1.1^\circ$ (longitude) with 32 vertical levels (maximum at 10 hPa). Each simulation runs for a period of 3 years.

\subsection{Target variable calculation}
Following \citeA{elsaesser2024}, we choose 33 target variables to emulate for the ModelE3 PPE. Seven of them are calculated based on global or regional model outputs (e.g., top of the atmosphere outgoing long wave radiation or Amazon rainfall), and the rest denote model performance scores (i.e., "metrics") which are functions of model outputs and satellite observations and their uncertainties. All outputs are described in \citeA{elsaesser2024}. The novelty in the formulation of these metrics is that the systematic error from different satellite products is taken into account in the metric calculation. We do not describe them in detail due to space limitations.

Three variables (i.e., \textit{T$\_$100hpa}, \textit{qv$\_$100hpa} and \textit{Trop$\_$Cyclone$\_$Count}) used in \citeA{elsaesser2024} are excluded here because the former two have very few non-zero values given the uncertainty in the observations for these variables exceeds the typical PPE-member simulated output (see Fig. 4e of \citeA{elsaesser2024}), while the last output variable is not continuous. Since some output variables from the ModelE3 PPE are not available in the CAM6 PPE, we do not have the equivalent number of target variables in the two PPEs. This does not affect any interpretations or arguments derived from our analyses. A brief description of the 33 target variables and whether they are calculated based on the CAM6 PPE are given in Table \ref{t_table1}.

\begin{table}[H]
    \caption{Variables of interest in this work. }
    \noindent\includegraphics[width=1.5\textwidth]{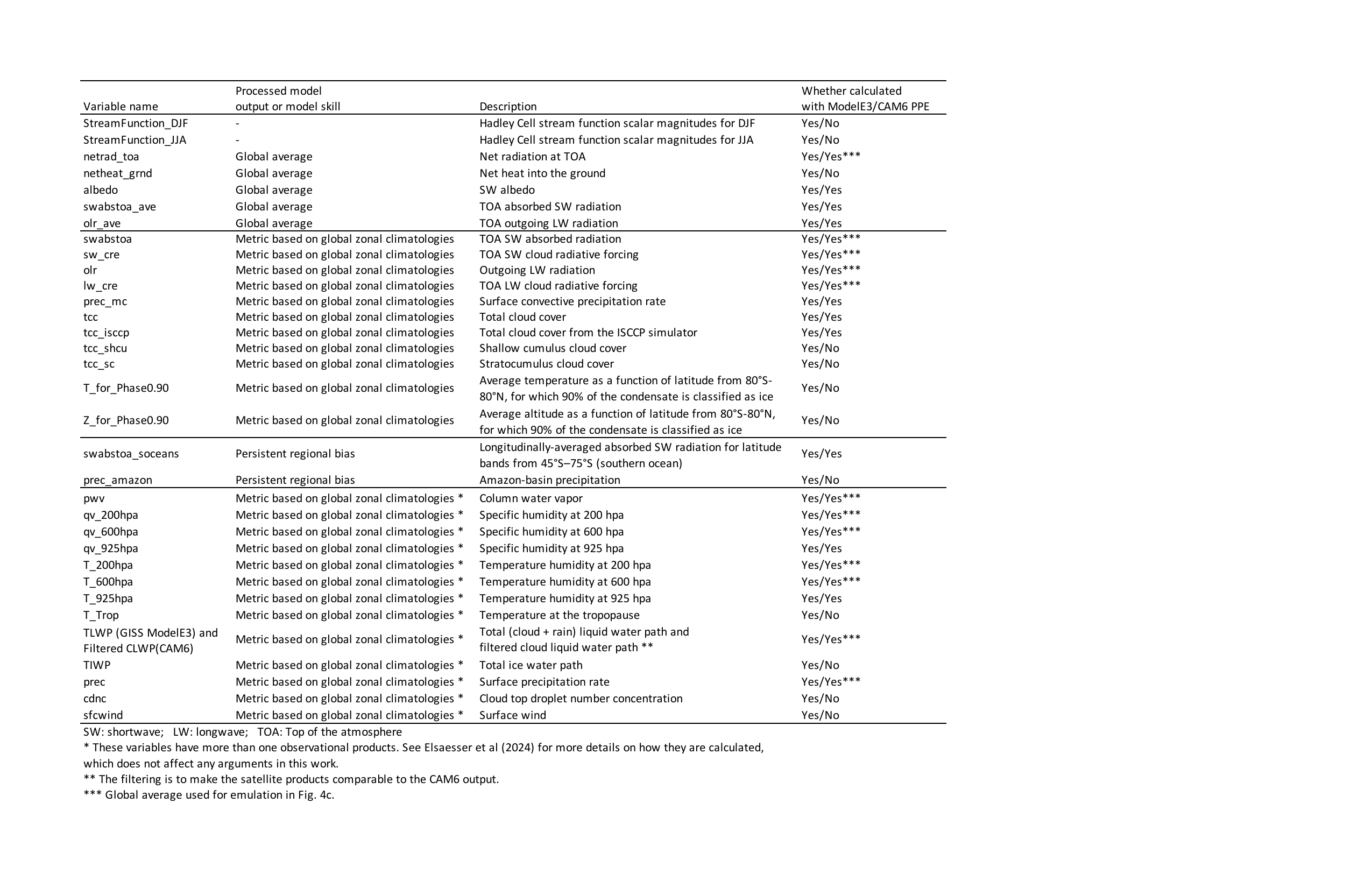}
    \label{t_table1}
\end{table}

\section{Method}
Climate model PPEs are commonly characterized by high dimensional parameter spaces (e.g., several tens of parameters are common) and limited numbers of ensemble members (e.g., in most cases less than 1000), implying sparsely-sampled parameter spaces. Owing to sparsity, when the target output variable is a complicated function of the parameters (e.g., non-linear and involving multi-parameter interactions), parameter -- output relationships may be difficult to detect, regardless of the emulation method used. This is analogous to drawing the coastline of a country given three points on a map. This motivates us to develop \textit{sage} that enables detecting relationships more robustly in a PPE.

\textit{sage} is a simplified technique that utilizes aspects of additive GPs \cite{duvenaud2011additive,cheng2019additive,lu2022additive}. It is made available on github (\url{https://github.com/yiqioyang/ppesage}). For a commonly used GP emulator that is non-additive, only one GP is trained and used for the prediction. Such a single GP is a function of all parameters. Conversely, for an additive GP emulator, its prediction is the sum of multiple GPs, with each GP being a function of different parameters and (or) parameter groups. In \textit{sage}, we do not utilize a formal additive GP (i.e., we do not estimate the hyperparameters) for the following reasons: (1) in practice, we find that estimating the hyperparameters of an additive GP could lead to overfitted results, though whether this occurs more generally depends on the methods used to estimate the hyperparameters in different GP-related libraries and the initial guess on the hyperparameters which is required as input by some libraries; and (2) more importantly, we find that the values of the hyperparameters would not greatly affect the \textit{sage} performance. This marks a key finding from this work, which we further assess later.

\subsection{Additive emulator}
\textit{sage} treats one climate model output variable at a time. We denote one scalar output as $z$ and the $D$-dimensional parameter space as $z = f(x_1, x_2, ..., x_D)$ with $x_1, x_2, ..., x_D$ being the model parameters.  For one target variable, our emulator decomposes $z$ into additive terms, with each a function of one, two or three parameters, as follows:
\begin{linenomath*}
\begin{equation}
    f(x_1, x_2, ..., x_D) \approx  
        \sum_{j=1}^{M_1} h_{1,j}(x_{I_{1,j}}) + 
	    \sum_{k=1}^{M_2} h_{2,k}(\boldsymbol{X_{2,k}}) + 
	    \sum_{l=1}^{M_3} h_{3,l}(\boldsymbol{X_{3, l}}),
    \label{eq_emu_a}
\end{equation}
\end{linenomath*}
where $h_{1,j}(x_{I_{1,j}})$, $h_{2,k}(\boldsymbol{X_{2,k}})$, and $h_{3,l}(\boldsymbol{X_{3, l}})$ correspond to functions of one, two, and three parameters, respectively, and $M_1$, $M_2$, and $M_3$ are their total numbers. $x_{I_{1,j}}$ is a parameter indexed by $I_{1,j}$ (an integer ranging from 1 to $D$). 
$\boldsymbol{X_{2,k}}$ and $\boldsymbol{X_{3, l}}$ correspond to a parameter pair and a parameter group of three, respectively. We could include additive terms that are functions of four or more parameters, but such an implementation is found to be unimportant in our subsequent analysis, and hence not adopted. For several target variables, \textit{sage} constructs their emulators separately following the general format of Eq. \ref{eq_emu_a}.

We set each term in Eq. \ref{eq_emu_a} as a GP. Different from formal additive GP and the more commonly-used GP that is not additive, we specify the hyperparameters of the GPs in \textit{sage} rather than estimating their values, and we only focus on the means of the GPs ($\mu^{GP}$), and ignore the corresponding estimated uncertainty. Eq. \ref{eq_emu_a} can be re-written as:
\begin{linenomath*}
\begin{equation}
    f(x_1, x_2, ..., x_D) \approx 
        \sum_{j=1}^{M_1} \mu_{1,j}^{GP}(x_{I_{1,j}} | \boldsymbol{\theta_1}) + 
	    \sum_{k=1}^{M_2} \mu_{2,k}^{GP}(\boldsymbol{X_{2,k}}| \boldsymbol{\theta_2}) + 
	    \sum_{l=1}^{M_3} \mu_{3,l}^{GP}(\boldsymbol{X_{3, l}}| \boldsymbol{\theta_3}),
     \label{eq_final}
\end{equation}
\end{linenomath*}
where $\mu_{1,j}^{GP}$, $\mu_{2,k}^{GP}$, and $\mu_{3,l}^{GP}$ correspond to the means of the one-, two-, and three-parameter GPs, respectively. $\boldsymbol{\theta_1}$, $\boldsymbol{\theta_2}$, and $\boldsymbol{\theta_3}$ are the fixed, specified GP hyperparameters.

We do not estimate the GP hyperparameters because we only focus on emulating the non-local, low-frequency relationship between the parameters and the output variable. Otherwise, the high-frequency variation, which could be overfitted by the emulator (due to the sparsity of the PPE), will likely be mistakenly emulated as signal rather than more appropriately treated as noise.

The above additive design resembles the polynomial-based emulation methods adopted in \citeA{williamson2013history} and \citeA{bellprat2016objective}. The difference is that the additive terms in previous works are polynomial functions of the parameters. In \textit{sage}, since each additive term is the mean of a GP, it does not require the step of estimating the coefficient values required in polynomial-based methods, and we do not assume the format (i.e., GPs are non-parametric emulators) of the emulator.

\subsubsection{Aspects of Gaussian Processes used in this method}
The theoretical aspect of GPs is not critical to understanding \textit{sage}. The role of the GPs in this method can be understood as a non-parametric interpolation method. See \citeA{rasmussen2003gaussian} for a thorough and detailed introduction of GPs. We consider the below description sufficient in the context of the design of \textit{sage}. The estimated mean of the GP is the weighted sum of the response (i.e., target variable in the present context) of the training data. The weights are determined by the covariance function which characterizes how similar or ``close'' a point of interest in the parameter space is relative to the others. We use the Mat\`ern covariance function, the most common format of covariance function in this method \cite{rasmussen2003gaussian}. The covariance function is defined by several hyperparameters. In particular, the range is a length scale that controls how ``far'' two points in the parameter space are considered to be correlated. With a greater range, the GP tends to model the low-frequency relationships existing in the data, because points that are ``far'' from the point of interest will be considered when making the prediction, leading to an overall smoother hyperplane. With a smaller range, more local and high-frequency fluctuation can be captured. The nugget to (signal) variance ratio can be understood as the noise to signal ratio. With these two hyperparameters specified, the GP hyperparameters of the covariance function are determined for a given set of training data (see Eq. 11 of \citeA{gu2018robustgasp}). We note here that the nugget (also known as noise level in some analyses) and (signal) variance are treated as two hyperparameters in some literature \cite{rasmussen2003gaussian}. A simple example is given in Appendix A to illustrate how the range and nugget-to-variance ratio affect the emulated mean of a GP. \textit{sage} is built from the R package ``RobustGaSP'' \cite{gu2018robustgasp}, but it can be easily written in python and based on other GP libraries due to its simplicity.

\subsection{Defining the workflow and diagnostics that enhance interpretability}
\begin{figure}[H]
    \noindent\includegraphics[width=\textwidth]{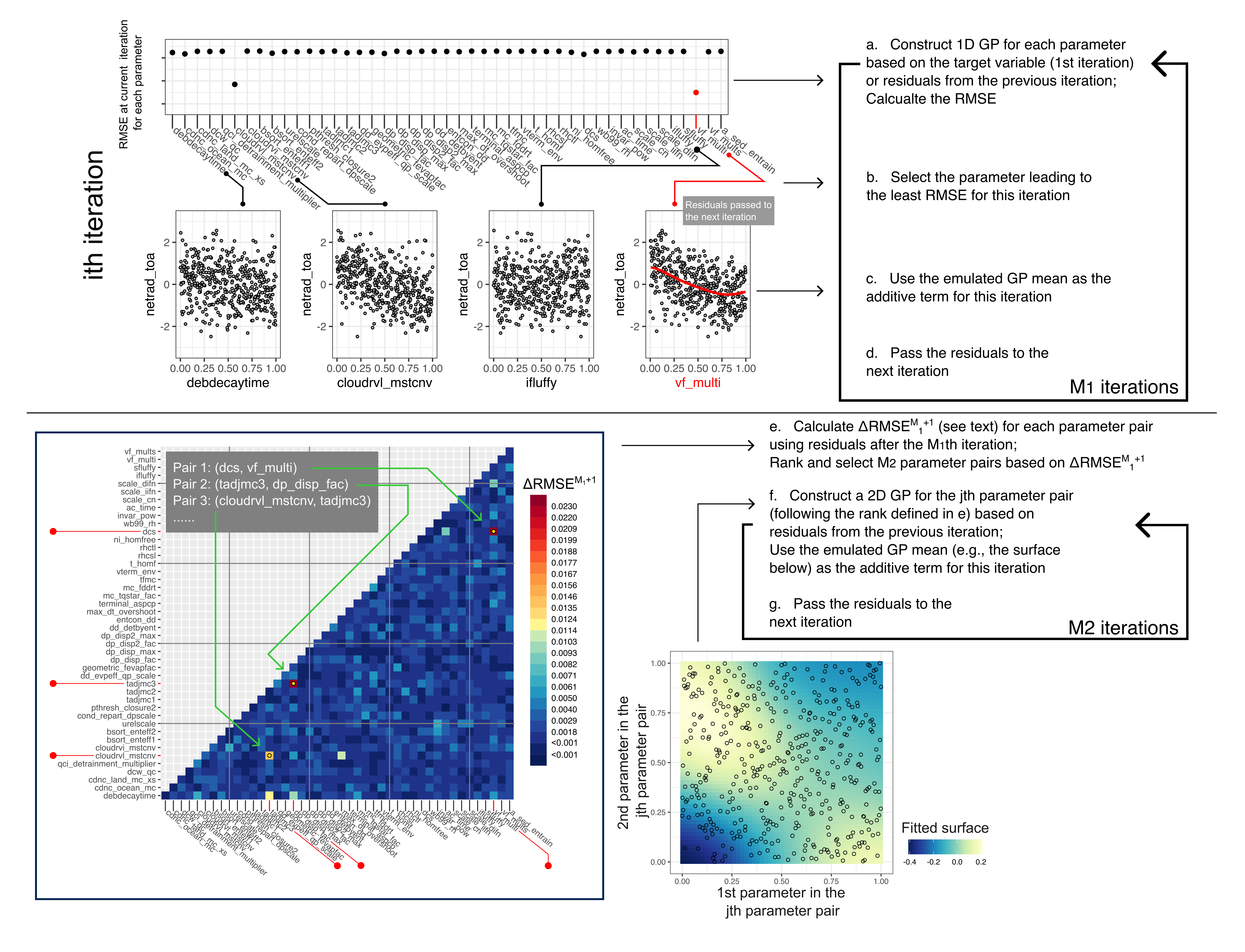}
    \caption{Workflow of \textit{sage} for individual parameters and parameter pairs. The variable \textit{netrad$\_$toa} in the ModelE3 PPE Default Test is used as an example here. }
    \label{f_flowchart}
\end{figure}

With the format of the emulator determined and the GP hyperparameters pre-specified, we must determine what parameters and parameter groups that $x_{I_{1,j}}$, $\boldsymbol{X_{2,k}}$, and $\boldsymbol{X_{3,l}}$ correspond to in Eq. \ref{eq_final} and the numbers of the additive terms (i.e., $M_1$, $M_2$, and $M_3$). Determination of the parameters and parameter groups can be visualized with the flow chart outlined in Fig. \ref{f_flowchart}. It can be generally viewed as a stepwise forward selection method (see more details on this topic in \citeA{draper1998applied}). A similar approach of method (parameter) selection has been adopted in the emulator of \citeA{williamson2013history,sexton2019finding}. The most distinct difference for \textit{sage} is its criterion to select the effective parameter interaction (Eq. \ref{eq_measure_pair}) which will be introduced later. We note here that the code to implement \textit{sage} is fully automated, but it is introduced in detail below with the hope that certain aspects of it can inform or be improved from future studies.

\begin{figure}[H]
    \noindent\includegraphics[width=0.8\textwidth]{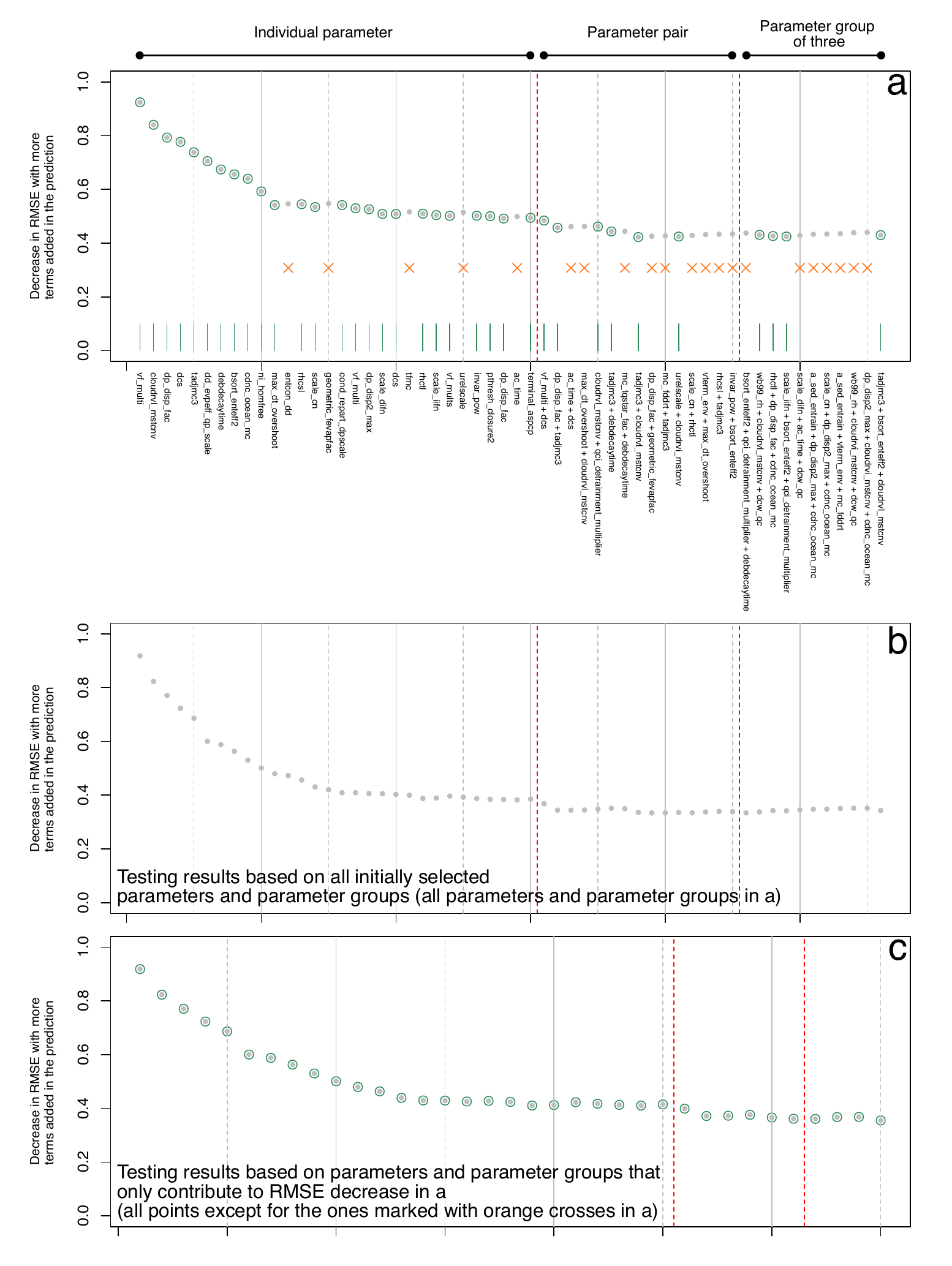}
    \caption{Example showing how \textit{sage} works using the example of \textit{netrad$\_$toa} from the ModelE3 PPE. The three figures are the output of \textit{sage}. In all three figures, the y-axis corresponds to the RMSE after more terms are included in the prediction. Its value decreases as more additive terms (x-axis, the corresponding parameters and parameter groups) are included in the prediction. The x-axes in b and c are not labeled for simplicity. a: applying the emulator trained on 80$\%$ of the training data to the remaining data. The points circled in green and marked by the green bars (and not marked by the orange crosses) contribute to the decrease of the RMSE, and are hence selected as the final parameters and parameter groups. 
    We train the complete training data with the originally selected (all points in a) and final (those that are not marked by the orange crosses in a) parameters and parameter groups, and apply them to the testing data. The results are presented in b and c, respectively. As a result, a and b share the exact same x-axis and number of points. There are fewer points in c compared to in a and b. }
    \label{f_validation_example}
\end{figure}

\textit{sage} begins with initial guesses on $M_1$, $M_2$, and $M_3$. 
Given $D$ parameters, setting the initial guesses as $M_1 = \text{int}(D \times 3/4)$, $M_2 = \text{int}(D/3) $, and $M_3 = \text{int}(D/4)$, where $\text{int}$ denotes taking the integer part of a number, are sufficient given our experience working with the two PPEs. \textit{sage} then uses 80$\%$ of the data to select $M_1$ single parameters, $M_2$ parameter pairs, and $M_3$ parameter groups of three (detailed below). It then applies the emulator trained on 80$\%$ of the data to the other 20$\%$ of the data, evaluates performance, and excludes the selected parameters and parameter groups that do not help with the emulation (i.e., the points that do not lead to a decrease in the RMSE in Fig. \ref{f_validation_example}a). The above process selects a sequence of parameters and parameter groups, and provides an emulator, although note that the emulator is trained only on 80$\%$ of the data. As the last step, \textit{sage} trains a final emulator based on the \textit{complete} dataset using the selected parameters and parameter groups from the previous step.

\subsubsection{Determining the parameters and parameter groups}
\textit{sage} focuses on determining single parameters ($x_{I_{1,j}}$) first, which is done iteratively and sequentially. In the first iteration, it aims at emulating the target variable, and in the following ones, it emulates residuals from the previous iteration. In each iteration, it (1) constructs a one-parameter GP for each of the $D$ parameters with the same pre-specified hyperparameters, (2) calculates and examines the RMSE from the $D$ GPs based on their mean predictions, and picks the parameter corresponding to the lowest RMSE as the selected parameter in this iteration, (3) uses this selected emulated GP mean as the additive term for this iteration and (4) passes the residuals from the fitting (based on the GP mean) to the next iteration. The above four steps correspond to Fig. \ref{f_flowchart}a-d in the flow chart. We allow a single parameter to be selected more than once in case that parameter could contribute to the fitting multiple times. In our experiments, the second time a parameter is selected either occasionally improves the emulator performance by a maximum of $10\%$ (of the overall RMSE) or does not introduce additional noise to the prediction (i.e., the corresponding additive term is nearly a constant zero). This feature is hence enabled.

After the $M_1$ iterations are complete for the single parameters, \textit{sage} moves on to determine the parameter pairs ($\boldsymbol{X_{2,k}}$). We will have $\binom{D}{2}$ parameter pairs. For \textit{each} of the parameter pairs $x_p$ and $x_q$, we construct three GPs for the residuals after the $M_1$th iteration (i.e., after the individual parameters are all selected): two one-parameter GPs for the two parameters respectively, and another two-parameter GP for $(x_p,x_q)$. The three GPs will lead to three RMSEs, denoted as $RMSE^{M_1+1}(p)$, $RMSE^{M_1+1}(q)$, and $RMSE^{M_1+1}(p,q)$, respectively. Denoting the $RMSE^{M_1}$ as the RMSE after the $M_1$th iteration (i.e., the RMSE of the prediction as the sum of $M_1$ additive terms), we use $\Delta RMSE^{M_1+1} (p, q)$ to evaluate and rank the additional improvement in the fitting if we consider the interaction of parameters pairs: 
\begin{linenomath*}
\begin{equation}
    \Delta RMSE^{M_1+1} (p, q) = (RMSE^{M_1} - RMSE^{M_1+1}(p,q)) - (2 * RMSE^{M_1} - RMSE^{M_1+1}(p) - RMSE^{M_1+1}(q)).
    \label{eq_measure_pair}
\end{equation}
\end{linenomath*}
The first two terms on the RHS (right hand side) reflect how much variability is explained by using the GP that considers $x_p$ and $x_q$ jointly, and the last three terms correspond to the variability explained by considering $x_p$ and $x_q$ independently. Greater $\Delta RMSE^{M_1+1} (p, q)$ indicates greater benefit in emulating the two parameters jointly. The calculation of Eq. \ref{eq_measure_pair} for all parameter pairs corresponds to step e in Fig. \ref{f_flowchart}.

We rank and select $M_2$ parameter pairs with the greatest $\Delta RMSE^{M_1+1} (p, q)$. \textit{sage} then sequentially emulates the residuals from the previous iteration, and passes the residuals from the fitting to the next iteration using each of the selected and ranked parameter pairs as the GP inputs (Fig. \ref{f_flowchart}f and g). A similar procedure is then implemented for parameter groups of three after a total of $M_2$ + $M_1$ iterations. The only difference is that we modify Eq. \ref{eq_measure_pair} for parameter groups of three such that it denotes the additional benefit of emulating three parameters jointly compared to emulating them independently.

We could select parameter pairs and parameter groups of three through iterations, as is done for the individual parameters. This is easy to implement in the code, although it would be computationally time-consuming, and thus is not adopted in the current version of the code. The complete training process leads to two products: (1) a sequence of several parameters, parameter pairs, and parameter groups of three and (2) the trained emulator for the target variable.

\subsection{Including the relationship between outputs}
The level of difficulty in emulating different target variables may vary. It is possible that an easier-to-emulate variable A could help emulate variable B that is more difficult to emulate. We can explore whether such relationships exist by including output variables that are easier to estimate as parameters, and apply \textit{sage} focusing on emulating the other variables. We pursue this in Section 6.3.

\section{Emulator performance}
As mentioned, we apply \textit{sage} to the ModelE3 and CAM6 PPEs independently. We evaluate its performance using randomly sampled ensemble members for training and validation and with varied GP hyperparameters. Its performance is compared with a NN emulator. How each parameter and parameter group contributes to the emulation is a key component of our workflow and analyses.

The NN used in this work is the ensemble average of four fully connected NNs with varying numbers of hidden layers (3-5), nodes (32-256), and activation functions (ReLU and leaky ReLU). The NN emulates all variables all at once. We choose to compare against a NN emulator over the conventional, non-additive GP since the NN slightly outperforms the latter (Appendix B).

\subsection{Default and random sampling tests}
We first define Default Tests for the two PPEs, respectively, for better visualization and easier examination of subsequent results. For the ModelE3 PPE, we use the first 1-451, 552-631, and 652-731 (a total of 611) ensemble members in the complete dataset to train the emulator, and set aside 140 for validation. For the CAM6 PPE, the first 210 ensemble members are selected for training, with 52 retained for validation. Note that the last 200 ensemble members (552-751) in the ModelE3 PPE are clustered around a region in the parameter space with greater likelihood of approximating the observations. We take 160 samples from them to ensure that the training data are representative of the complete PPE. Both Default Tests retain 18-19$\%$ of the ensemble members for validation.  In addition to the Default Tests, we also randomly draw ensemble members for training and validation from both PPEs ten times (another ten random sampling tests). The numbers of training and validation members are identical to those in the respective Default Tests. These random sampling tests are implemented to detect how the ensemble members used for training and validation affect the emulator performance. Throughout this work, the same 11 sets (a Default Test plus ten random sampling tests) of training and validation datasets are used for each PPE for consistency unless specified otherwise. In working with the ModelE3 PPE, the parameter sequences are based on the training samples that are within the first 551 ensemble members such that the parameter sequences are obtained based on data that are evenly distributed in the parameter space. These parameter sequences are then coupled with the complete training data to train the final emulators.

We apply both \textit{sage} and NN to the 11 sets to evaluate and compare their performance for each PPE. For both methods, different emulators are trained in each test. The GP hyperparameters are fixed and provided in Table \ref{t_hyper_parameters}.  These will be varied later, as explained. All parameters are normalized to range from zero to one, and target variables are standardized to be zero-mean with a standard deviation of one. The hyperparameters ``ranges'' specified in Table \ref{t_hyper_parameters} ensure that they are at least 30$\%$ (i.e., range for 3-D GP in Set 5; $\sqrt{0.3^2 + 0.3^2 + 0.3^2}/\sqrt{1^2 + 1^2 + 1^2}$) of the maximum distance ($\sqrt{1^2 + 1^2 + 1^2}$) of the parameter spaces, which ensures that we emulate the non-local, low-frequency relationship between the target variables and parameters.

\begin{table}
  \centering
  \caption{Specified GP hyperparameters}
  \begin{tabular}{|c|c|c|c|c|}
    \hline
    Set name & Range for 1-D GP & Range for 2-D GP & Range for 3-D GP & Nugget to variance Ratio \\
    \hline
    Default Test & 0.60 & $\sqrt{0.5^2 + 0.5^2}$ & $\sqrt{0.4^2 + 0.4^2 + 0.4^2} $ & 2.00 \\
    Set 1   & 0.60 & $\sqrt{0.5^2 + 0.5^2}$ & $\sqrt{0.4^2 + 0.4^2 + 0.4^2} $ & 4.00 \\
    Set 2   & 0.60 & $\sqrt{0.5^2 + 0.5^2}$ & $\sqrt{0.4^2 + 0.4^2 + 0.4^2} $ & 1.00 \\
    Set 3   & 0.80 & $\sqrt{0.6^2 + 0.6^2}$ & $\sqrt{0.4^2 + 0.4^2 + 0.4^2} $ & 2.00 \\
    Set 4   & 1.00 & $\sqrt{0.8^2 + 0.8^2}$ & $\sqrt{0.6^2 + 0.6^2 + 0.6^2} $ & 2.00 \\
    Set 5   & 0.50 & $\sqrt{0.4^2 + 0.4^2}$ & $\sqrt{0.3^2 + 0.3^2 + 0.3^2} $ & 2.00 \\
    \hline
  \end{tabular}
  \label{t_hyper_parameters}
\end{table}

\section{Emulator performance}
\begin{figure}[H]
    \noindent\includegraphics[width=\textwidth]{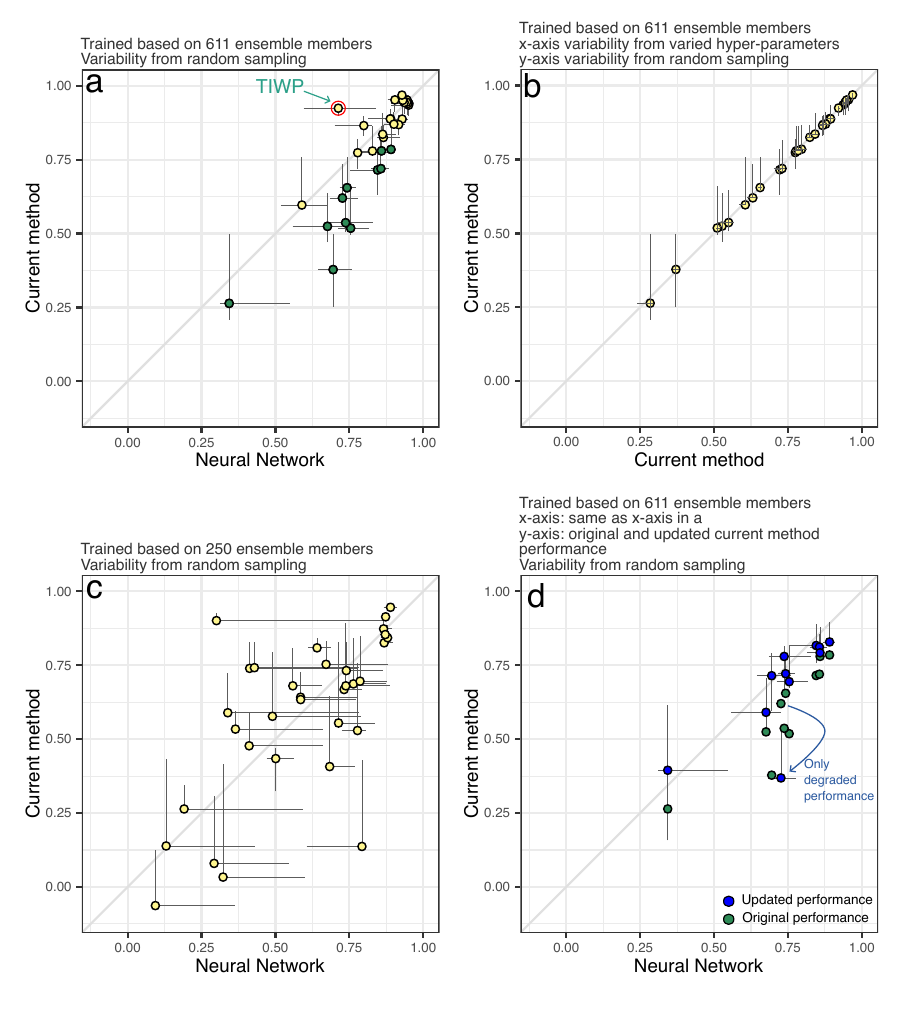}
    \caption{R-squares and their variability from different tests with the ModelE3 PPE using \textit{sage} and the NN. 
    a and c: comparison between \textit{sage} and NN given 611 and 250 ensemble members for training. Green points in a correspond to variables that NN is better at emulating. The points are marked with different colors in a as the green points will be used for additional analysis (shown in d). Since no additional analysis is done for points in c, we do not label them with different colors.  
    b: R-square variability using \textit{sage} from random sampling (y-axis) and from varied GP hyperparameters (x-axis) with 611 ensemble members for training. d: updated performance (blue points) of \textit{sage} (see text for details) compared with its original performance (green points) and NN trained with 611 ensemble members for the variables that the NN is better at predicting (green points in a). All points are plotted based on the Default Test (a, b, and d) or the test that uses the first 250 ensemble members (c) for training. All vertical and horizontal bars denote variability from random sampling except for the horizontal bars in b which denote the variability from varied GP hyperparameters. }
    \label{f_giss_rsq}
\end{figure}

\begin{figure}[H]
    \noindent\includegraphics[width=\textwidth]{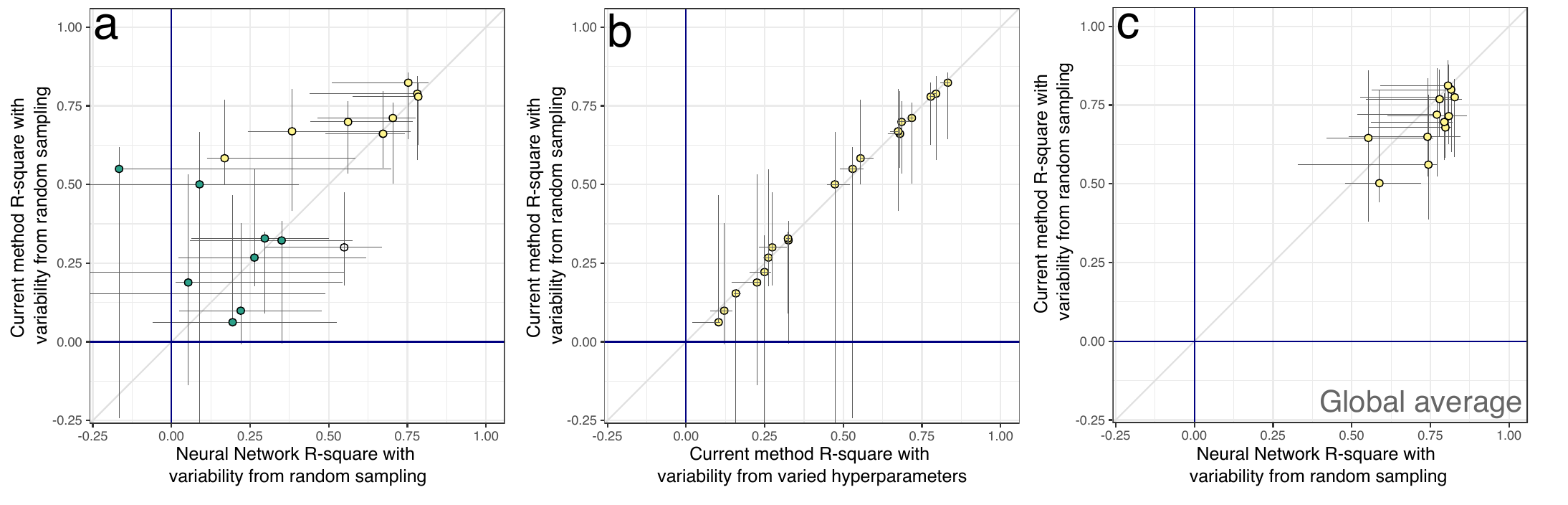}
    \caption{R-squares and their variability with the CAM6 PPE using \textit{sage} and NN. a: R-squares from using the two methods. The points are plotted based on the Default Test results, and the bars denote the variability from random sampling. Variables with low performance (e.g., overall lower R-square or significant variability in R-square) using \textit{sage} marked in green; b: R-squares of \textit{sage} with the variability from random sampling (vertical bars) and from varied GP hyperparameters (horizontal bars). Points in a and b are plotted based on the Default Test. More points are shown in b than a because some variables have their R-squares outside the extent in a using NN; c: the same as a except that the target variables are all global averages. These variables are marked in Table \ref{t_table1}.}
    \label{f_cam_rsq}
\end{figure}

The resultant R-squares of the two PPEs and their variability from using the two methods are compared in Figs. \ref{f_giss_rsq}a and \ref{f_cam_rsq}a. The performance of the two methods is comparable for the ModelE3 PPE with the NN exhibiting better performance for the variables that are relatively difficult to emulate (i.e., those with lower R-square for both methods). Among the 33 target variables, \textit{sage} outperforms the NN in emulating variables
\textit{TIWP} (Total Ice Water Path; highlighted in Fig. \ref{f_giss_rsq}a) and \textit{tcc$\_$sc} (stratocumulus cloud cover) in both the Default Test R-squares (absolute R-square differences between the two methods: 0.21 and 0.07) and its variability from random sampling. Twenty variables have their R-square differences between the two methods within the range of -0.05 and 0.05 in the Default Test. The corresponding R-square variability is also comparable. For the other 11 variables, marked green in Fig. \ref{f_giss_rsq}a, the R-square differences between the two methods exceeds 0.05 (two variables below 0.10; five in the range of 0.10-0.20; and four in the range of 0.20-0.32). The NN outperforms \textit{sage} in emulating these. 

Both methods exhibit degraded performance in emulating the CAM6 PPE relative to the ModelE3 PPE. For CAM6, \textit{sage} has an overall improved performance relative to the NN (Fig. \ref{f_cam_rsq}a), and both are characterized by large variability in R-square in the random sampling tests. Among the 21 variables emulated, there are 6 (there is a pair of variables overlapping in Fig. \ref{f_cam_rsq}a) variables that both methods emulate relatively and comparatively well: their R-squares in the Default Test are above 0.625 using both methods with similar variability from random sampling. \textit{sage} outperforms the NN in emulating another three variables with their corresponding R-squares being 0.58, 0.67, and 0.70 using the former in the Default Test. Their R-square variability is comparable or slightly better when \textit{sage} is used. These variables (6 + 3) are marked yellow in Fig. \ref{f_cam_rsq}a. The NN outperforms \textit{sage} in emulating the variable \textit{T$\_$925hpa} (empty circle in Fig. \ref{f_cam_rsq}a). Neither \textit{sage} nor the NN performs well in emulating the variables colored green in Fig. \ref{f_cam_rsq}a. They have either low or even negative R-squares in the Default Test or extreme R-square variability from the random sampling tests. (The negative R-squares arise when the emulator introduces additional bias and noise to the prediction, e.g., for a set of points on the line $y = x$, an emulator in the form of $y_{emu} = -x$ could lead to negative R-squares). Another three variables have their R-squares or variability outside the extent of Fig. \ref{f_cam_rsq}a and are not shown.

Considering the two PPEs, the two methods perform differently not only because the models, parameter ranges and samples are different, but because some target variables in both PPEs are skill score metrics rather than model output climatologies (Table \ref{t_table1}). We illustrate this in an additional analysis of the CAM6 PPE in Section 7.2.

\subsection{Varied hyperparameters}
Another five experiments are performed using the training and validation datasets in the Default Tests of the two PPEs. Different GP hyperparameters defined in Table \ref{t_hyper_parameters} (Sets 1-5) are used here, and the same parameter sequences as the Default Tests are used. The corresponding R-square variability is compared with that from the random sampling tests using \textit{sage} (Fig. \ref{f_giss_rsq}b and Fig. \ref{f_cam_rsq}b). The R-square variability from the varied hyperparameters is much smaller than that from the random sampling tests for both PPEs, showing that the GP hyperparameters having very limited impact on the emulator performance.

\subsection{Contributions from individual additive terms}
\begin{figure}[H]
    \noindent\includegraphics[width=\textwidth]{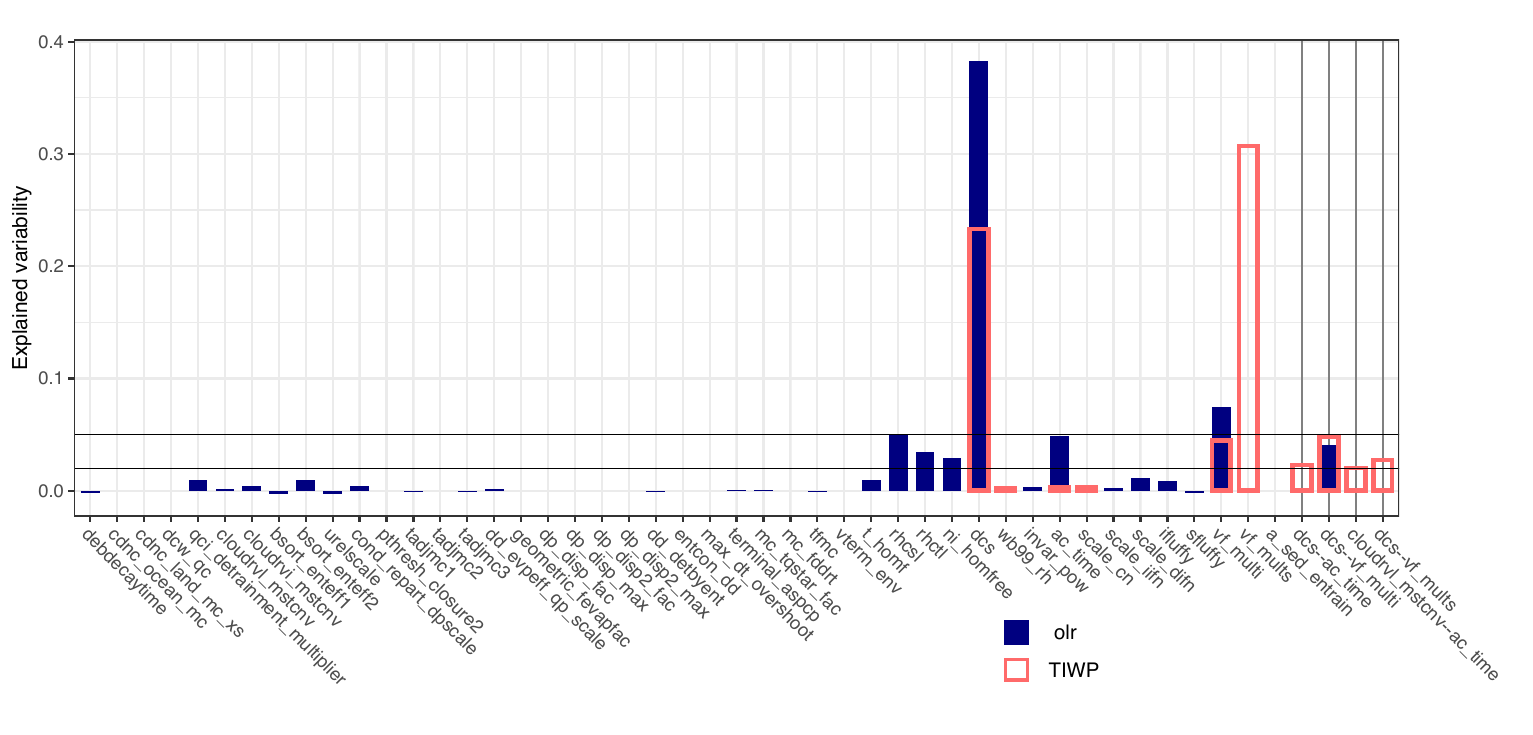}
    \caption{Explained variability for all individual parameters plus parameter groups with explained variability greater than 0.02 (marked by vertical lines) for variables ``olr'' and ``TIWP''. Calculated based on the ModelE3 PPE Default Test. }
    \label{f_giss_rmse_olr_tiwp}
\end{figure}

Since the emulator prediction is additive, we are able to examine how each parameter or parameter group contributes to the normalized RMSE decrease (normalized for the testing data) during the training or validation. The latter is preferred to ensure robustness. 
We call the decrease in the RMSE ``explained variability'' in the following text (Fig. \ref{f_validation_example}). Examples for target variables \textit{olr} (outgoing long wave radiation) and \textit{TIWP} in the ModelE3 PPE Default Test are presented in Fig. \ref{f_giss_rmse_olr_tiwp}. (We note here that \textit{sage} automatically outputs the explained variability from each parameter and parameter group.)

\begin{figure}[H]
    \noindent\includegraphics[width=\textwidth]{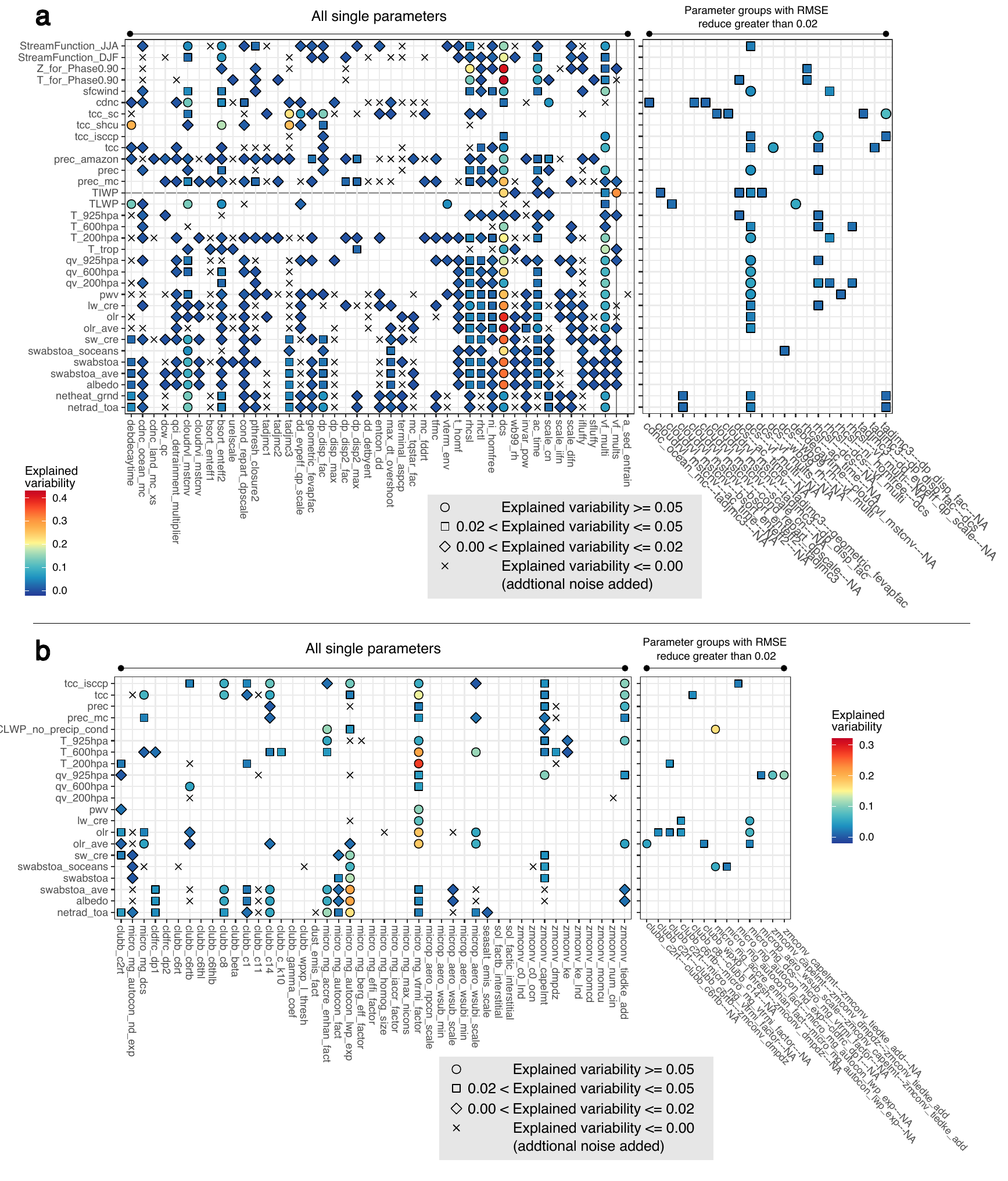}
    \caption{Explained variability from all individual parameters plus parameter groups with explained variability greater than 0.02 in the Default Tests of the ModelE3 PPE (a) and CAM6 PPE (b). The crosses denote the parameters that introduce additional noise in the prediction.}
    \label{f_both_rmse_grid}
\end{figure}

We calculate the explained variability for all variables based on the Default Tests of the two PPEs (Fig. \ref{f_both_rmse_grid}). For parameter groups, only those with explained variability greater than 0.02 are presented in Fig. \ref{f_both_rmse_grid} to avoid redundancy in the visualization. The crosses in Fig. \ref{f_both_rmse_grid} denote the individual parameters that contribute negatively to the prediction (negative explained variability). They are selected during the training process but shown to degrade the prediction in the Default Tests. Their total impact on the emulator performance is different for the two PPEs, which will be examined in Section 5.3

The most noticeable features shown in Fig. \ref{f_both_rmse_grid} are briefly summarized here. For the ModelE3 PPE, the parameter \textit{dcs} (threshold diameter for autoconversion of cloud ice to snow) contributes the greatest to the explained variability of most target variables. Parameter \textit{vf$\_$multi} (multiplier on coefficient for cloud ice fall speed) helps greatly with the emulation prediction, but the explained variability from it is smaller relative to \textit{dcs}. As a pair, the two parameters also contribute jointly to the emulation of several target variables (e.g., variable \textit{pwv}, column water vapor).

For the CAM6 PPE, there are also two parameters, namely 
\textit{micro$\_$mg$\_$autocon$\_$lwp$\_$exp} (KK2000 LWP exponent, see more details in \citeA{gettelman2015advanced}) and \textit{micro$\_$mg$\_$vtrmi$\_$factor} (ice fall speed scaling), that dominantly contribute to the emulator prediction as individual parameters. Parameter groups do contribute to the prediction of some variables but with explained variability much smaller compared to the parameter groups in the ModelE3 PPE. 
%%%%

\subsubsection{The unexpected success in emulating \textit{TIWP} for the ModelE3 PPE and its implication}
\textit{sage} greatly outperforms the NN in emulating the ModelE3 PPE \textit{TIWP} (highlighted in Fig. \ref{f_giss_rsq}a). This markedly improved emulation was unexpected. Successful emulation is dominated by roughly 6 parameters (the parameters that are crossed by the horizontal line in Fig. \ref{f_both_rmse_grid}a), suggesting that this variable is not difficult to emulate by \textit{sage}. Why such an improvement? Parameter \textit{vf$\_$mults} (multiplier on coefficient for snow fall speed) contributes the most to the emulation of \textit{TIWP}, and it contributes very little to all other variables. The relationship between \textit{TIWP} and \textit{vf$\_$mults} is effectively independent from the remaining target variables and parameters. Because of this independence, we suspect that for a NN with 33 variables to emulate and with only 611 ensemble members to train from, it is unable to isolate the single parameter-variable relationship (i.e., \textit{TIWP} and \textit{vf$\_$mults}) well compared with the multi-variable-parameter relationship.

We delve deeper by constructing another two NNs focusing solely on ``TIWP''. The first one utilizes all 45 parameters as inputs and the second one utilizes the six parameters revealed as most important for \textit{TIWP} by our new method (i.e., the parameters that are crossed by the horizontal line in Fig. \ref{f_both_rmse_grid}a). The R-square, based on training and testing ensemble members from the Default Test, from the first NN varies greatly, ranging from 0.65 to 0.85 (in most cases around 0.69-0.75) in different training sessions; the second NN leads to R-square with limited variability, ranging from 0.86 to 0.93, a range comparable to the performance of \textit{sage}. The above results are invariant to different NN structures and setups (e.g., numbers of hidden layers, activation function, epochs).

The above experiments reveal important implications on emulator design for climate model PPEs. They show that the performance of the NN could be improved if two or more emulators are constructed to emulate separate groups of target variables. They are also consistent with our hypothesis that the better performance of \textit{sage} on \textit{TIWP} is related to its stand-alone relationship with \textit{vf$\_$mults}, as once given the correct parameters selected by \textit{sage}, the performance of NN improves greatly. The above arguments are important as similar scenarios might exist for other PPEs or other emulator methods. To our knowledge, they are not yet pointed out in previous studies. The above arguments cannot be directly derived from a NN, showcasing the importance of analysis on climate model PPEs, which should be distinguished from analyses of surrogate PPEs based on emulators that might produce more error-laden outputs for some variables.

\subsection{Cumulative contributions from additive terms}
\begin{figure}[H]
    \noindent\includegraphics[width=0.7\textwidth]{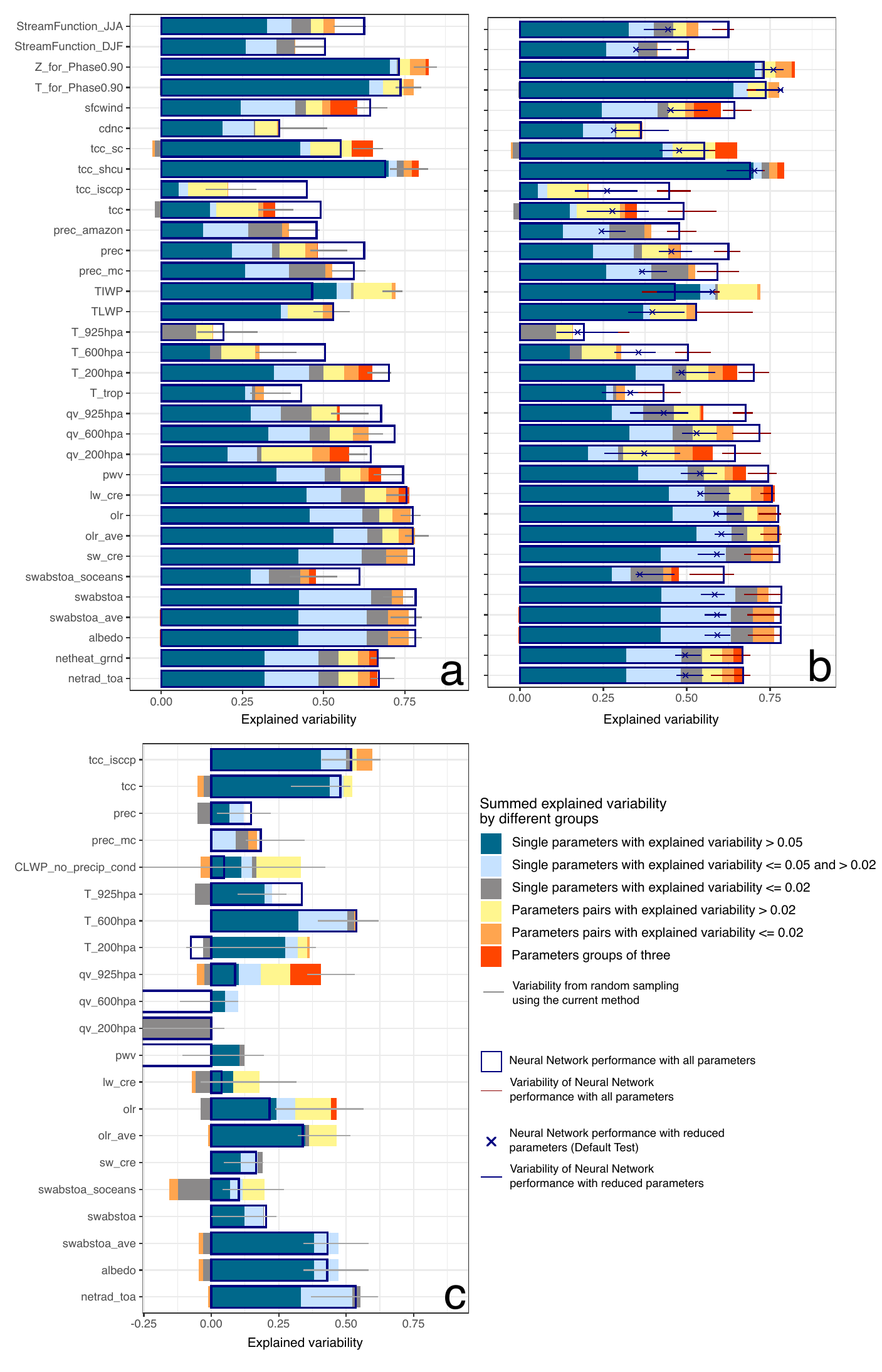}
    \caption{Explained variability (x-axis) grouped and summed based on different criteria in the Default Tests of the ModelE3 PPE (a and b) and CAM6 PPE (c) denoted as different colored bars. The total variability explained from the NN is marked using the dark blue empty boxes. The total variability from random sampling using \textit{sage} is plotted as horizontal lines in a and c. b is identical to a except for the following: the crosses denote the total explained variability from a NN trained on 13 selected sensitive parameters (see text for more details). It is based on the Default Test training and testing data. The red and blue bars in b denote the variability of the total variability explained using the NN due to random sampling based on all and the 13 sensitive parameters, respectively.}
    \label{f_both_rmse_summary}
\end{figure}

We group and sum the explained variability from each additive term based on the number of parameters it corresponds to (i.e., 1,2, and 3) and the value of the explained variability in the Default Tests of the two PPEs. We choose 0.05 and 0.02 as the thresholds for the latter criterion. The group that contributes to less than 0.02 RMSE includes the terms with negative contribution (i.e., crosses in Fig. \ref{f_both_rmse_grid}). The classification based on the value of the explained variability is not done for parameter groups of three as their total contribution is already small in most cases. The results are presented in Fig. \ref{f_both_rmse_summary} for the two PPEs.

For the ModelE3 PPE (Fig. \ref{f_both_rmse_summary}a), the variability in the total (summed) explained variability from random sampling is relatively small for most target variables, allowing us to draw the following points with confidence (see Appendix C, which further confirms the robustness of the points below). These points are listed with brevity but their importance should not be neglected: (1) most of the explained variability is from the contributions of individual parameters; (2) surprisingly, the cumulative (as \textit{sage} is additive) effect of parameters that individually contribute to less than 0.05 of the explained variability is not negligible (which we separate into two categories: between 0.05 and 0.02, and below 0.02); (3) The explained variability from parameter groups is relatively small, but cannot be neglected (e.g., 
\textit{TIWP}); (4) the contribution from parameter pairs is in most cases greater than that from parameter groups of three. In most cases (with two exceptions with negative gray bars in Fig. \ref{f_both_rmse_summary}a), the group with individual explained variability less than 0.02 improves the emulator performance, confirming that the overall impact of the additive terms with negative explained variability (crosses marked in Fig. \ref{f_both_rmse_grid}a) is limited for the ModelE3 PPE.

The second point listed above deserves additional attention, as it is related to how we should interpret the sensitivity of climate model parameters. This can be seen with the variable \textit{olr} in Fig. \ref{f_giss_rmse_olr_tiwp}. Five terms (four as individual parameters and one as parameter pair) have their explained variability in the range of 0.02-0.05 with a total sum of more than 0.15, an amount that should not be neglected in the emulator construction or sensitivity analysis. The cumulative effect of these individually less important parameters will be further studied with NN in Section 6.2.

Analysis of the ModelE3 PPE also shows that the parameter importance and sensitivity can be underestimated if we examine it without considering parameter interaction. We use variable \textit{TIWP} as an example to illustrate this (Fig. \ref{f_giss_rmse_olr_tiwp}). For this variable, parameter \textit{dcs} is not only important individually, it also pairs with another three parameters. The summed explained variability from these three pairs is greater than 0.09 (two above 0.02 and one around 0.05). This is not negligible as the total explained variability for this variable is 0.72 in the Default Test.

Results (Fig. \ref{f_both_rmse_summary}c) for the CAM6 PPE show that most effective and robust explained variability is from individual parameters that contribute to more than 0.05 of the explained variability. The cumulative effect of other individual parameters and parameter groups is in most cases relatively small or comparable to the total explained variability from the random sampling tests. There are not that many parameters or parameter groups that contribute to the degradation of the fitting (i.e., fewer crosses in Fig. \ref{f_both_rmse_grid}b; compared to the case with the ModelE3 PPE), but their individual negative impact is large (i.e., the wide negative bars in Fig. \ref{f_both_rmse_summary}c). The performance of both methods is low, suggesting the greater difficulty in emulating the CAM6 PPE. The variability in the total explained variability due to random sampling is relatively large, preventing us from making further interpretations.

Since the performance of \textit{sage} varies greatly for the two PPEs, they are analyzed separately below. For the ModelE3 PPE, we focus on what insights can be obtained from working with \textit{sage} and comparing it with NN. For the CAM6 PPE, we focus on measures to improve the emulator performance.

\section{Analysis on the ModelE3 PPE}
We focus on three aspects of the ModelE3 PPE: how the performance of \textit{sage} or NN changes in response to reduced (1) number of training ensemble members and (2) number of parameters, and (3) whether the variables that \textit{sage} emulates well could inform the emulation of those that are difficult to predict.

\subsection{Reduced number of ensemble members}
We used 250 ModelE ensemble members to train \textit{sage} and the NN, and evaluate their performance using the remaining ensemble members. The first 250 ensemble members plus another ten randomly sampled datasets of same size are used for training and the rest for validation, which amounts to 11 tests. The GP hyperparameters of the Default Test are also used here. The R-square and its variability from the 11 tests using \textit{sage} and NN are presented in Fig. \ref{f_giss_rsq}c. The points are evenly split by the one-to-one line, suggesting comparable performance from the two methods. For some variables, the variability from random sampling is much smaller when \textit{sage} is used, indicating its more robust performance given limited training data.

\begin{figure}[H]
    \noindent\includegraphics[width=0.7\textwidth]{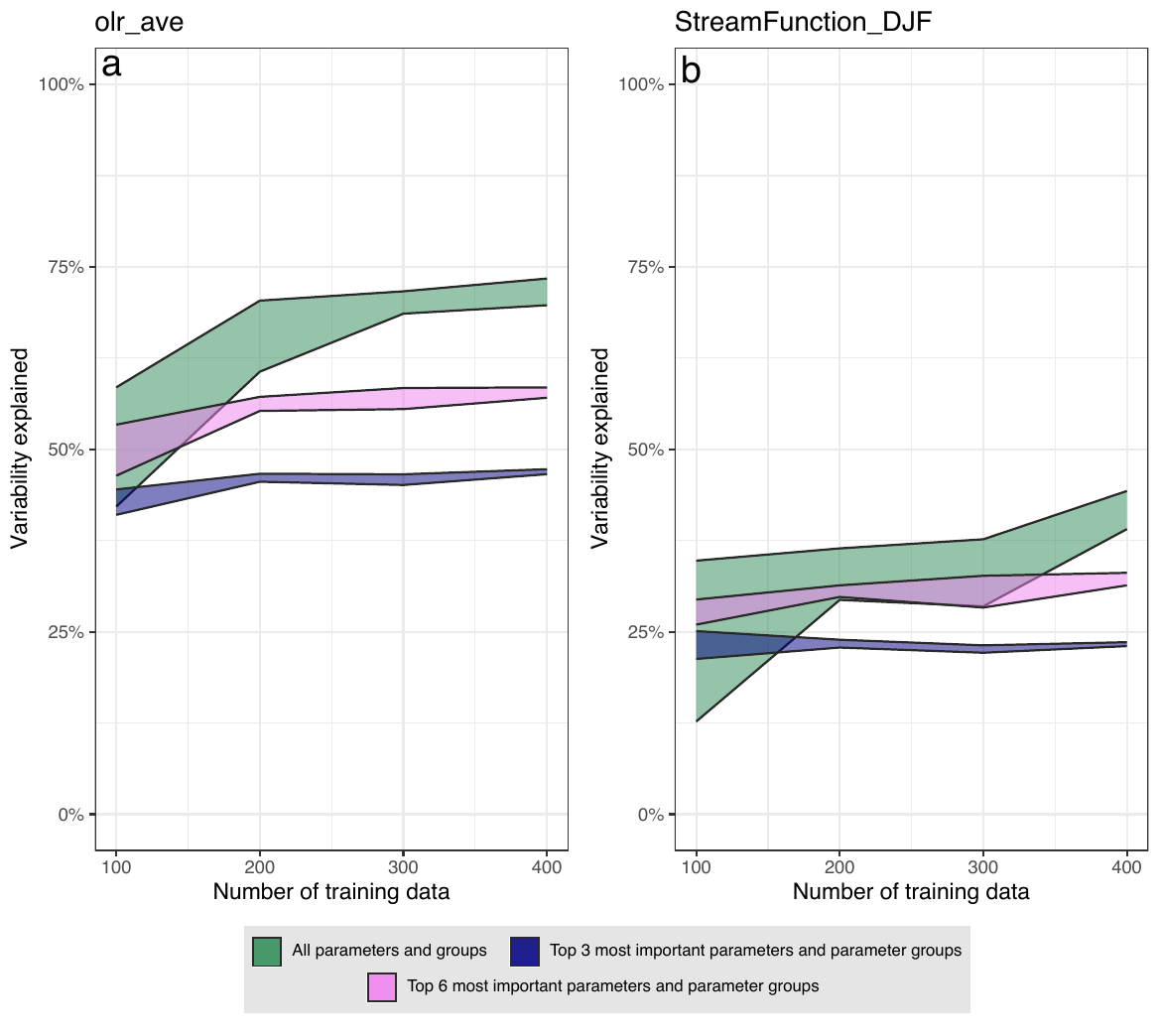}
    \caption{Total variability explained by all, the first three, and the first six most important parameters and parameter groups for the variables \textit{olr$\_$ave} (a) and \text{StreamFunction$\_$DJF} (b) for the ModelE3 PPE given different numbers of ensemble members for training. The variability is calculated based on applying emulators trained on randomly sampled ensemble members (from the first 451 ensemble members) to the same 452-551 ensemble members.  }
    \label{f_size_matter}
\end{figure}

We select variables \textit{olr$\_$ave} and \textit{StreamFunction$\_$DJF} for more detailed analysis. They are chosen since the \textit{sage} performance on the two variables are greatly different (comparable performance for the former using the two methods and lower performance for the latter using \textit{sage}; Fig. \ref{f_both_rmse_summary}). We randomly select 100, 200, 300, and 400 ensemble members from the first 451 ensemble members for training and then apply the trained emulators to the 452-551 ensemble members for validation and the calculation of the explained variability. This process is repeated five times (but the target is always the 452-551 ensemble members) for each number of ensemble members to produce ranges of variability explained. We do not consider the other ensemble members (552-751 ensemble members) as they are not randomly distributed in the parameter space, which might be a factor affecting the variability explained by \textit{sage}. The total variability together with the variability explained by the first three and six most important parameters and parameter groups are presented in Fig. \ref{f_size_matter}.

For \textit{olr$\_$ave}, the total explained variability increases with the number of training ensemble members, and plateaued after training on 300 ensemble members. For this variable, if we already have 300 ensemble members, the expected emulator improvement upon sampling another 100 ensemble members is negligible. In contrast for \textit{StreamFunction$\_$DJF}, as we increase from 300 to 400 members, the variability explained by all parameters and parameter groups increases by nearly 15 percent. The above results suggest that the emulator performance can be demonstrably improved by increasing the number of training ensemble members, although this conclusion is output-variable specific. It suggests an iterative approach to sampling, weighing the quantitative increase in variability explained against the cost of additional samples.

What the two experiments share in common is that the variability explained by the first three and six most important parameters does not change greatly with the number of training ensemble members. This is not surprising, given that most of the variability explained tends to be from individual parameters (Fig. \ref{f_both_rmse_grid}), and such independent and individual effects of these parameters do not need that many (in this case, 300 or 400) ensemble members for the emulators to capture.

For \textit{olr$\_$ave}, the increase in explained variability from 200 to 300 ensemble members is mostly from individually less important parameters and parameter groups (this is suggested in Fig. \ref{f_both_rmse_summary} and the section below). Fig. \ref{f_size_matter}a shows that the increase in explained variability does not change linearly with the number of training ensemble members. Quantitative analyses such as those presented in this section could be of use when deciding PPE design and member count.

\subsection{Reduced parameter space}
As shown earlier, parameters that do not contribute greatly to the emulator performance individually (i.e., those with explained variability below 0.05) could together have a non-negligible and cumulative impact on the emulator performance (Fig. \ref{f_both_rmse_summary}a). This is demonstrated using \textit{sage}, but might not be valid for other emulators. To test its universality, we keep the 13 sensitive parameters that individually contribute to more than 0.05 of explained variability to at least one target variable in the ModelE3 PPE Default Test, exclude the other parameters, and implement the Default Test plus the corresponding 10 random sampling tests using the NN. The same procedure is not performed for \textit{sage} as it is additive, and the results are hence already presented in Fig. \ref{f_both_rmse_summary}a.

The total explained variability and its variability using NN with the complete and reduced parameter spaces are compared in Fig. \ref{f_both_rmse_summary}b. For a variable whose explained variability from individual parameters is low but from combined parameters is large (i.e., those with long light blue and gray bars in Fig. \ref{f_both_rmse_summary}a and b ; e.g., \textit{albedo}), the degradation of the NN performance is consistently large with the reduced parameter space. The NN performance remains unchanged for the variables that are dominated by individually more important parameters (i.e., those with long dark blue bars and short light blue and gray bars in Fig. \ref{f_both_rmse_summary}a and b; e.g., \textit{tcc$\_$shcu}, shallow cumulus cloud cover). The cumulative effect of individually less important parameters is thus important to the NN as well, suggesting that its importance is not method-specific.

\subsection{Relationship between variables}
The \textit{TIWP} emulation case discussed earlier suggests it is not necessarily optimal to emulate all variables simultaneously. At the same time, it is intuitive that some target variables can be related to one another (arising from underlying physics), and leveraging such  relationships could improve emulator performance.

We explore this in the ModelE3 PPE, focusing on the variables that are difficult to emulate by \textit{sage} (i.e., the points marked green in Fig. \ref{f_giss_rsq}a). We select the variables with R-square greater than 0.85 in the Default Test using \textit{sage} as the easy-to-emulate variables, with the expectation that they could help emulation of the ``difficult'' variables.

During the training, we include the ``easy'' variables as parameters, and obtain a parameter sequence for each ``difficult'' variable based on the training dataset. The parameter sequences therefore have the original parameters and the ``easy'' variables. The ``easy'' variables are provided by the original emulators (i.e., the ones trained using the original 45 parameters). The 11 training datasets (i.e., the Default Test and the corresponding 10 random sampling tests) with 611 ensemble members are used for training here.

The original (i.e., trained with the original parameters) and updated (i.e., trained with the inclusion of the ``easy'' variables as parameters) R-squares of the ``difficult'' variables are plotted and compared in Fig. \ref{f_giss_rsq}d.  The R-squares increase for all ``difficult''  target variables with one exception. There are some variables with significant increase in the updated R-squares in all 11 tests, suggesting robust improvement in the emulator performance. 
The experiment results shown in Fig. \ref{f_giss_rsq}d suggest that target variables that are related should be grouped together to optimize the emulator performance.

The improved performance is related to how the target variable is defined. In the original emulator experiments, some target variables are defined to be model penalty scores (e.g., output of 1 and -1 judged relative to an observation of 0 is equivalently 1 to the model score), which prohibits detectable relationships that could help with the emulator performance. This increases the difficulty in emulating such model scores if the emulator is not informed of the relationship between target variables, i.e., how the original emulators are trained in the Default Test. A simple example is given in Appendix E to illustrate this. The explained variability from the additive terms for the ``difficult'' variables in this experiment together with biplots of selected parameters and variables are given in Appendix D. It shows that most of the variability is explained by those ``easy'' variables used as parameters here. This improved performance is not presented as the main results in this work because a subjective PPE-specific decision needs to be made on the definition of ``easy'' and ``difficult'' variables.

\section{Improving the performance of the CAM6 PPE emulator}

\subsection{Factors affecting the emulator performance}

\begin{figure}[H]
    \noindent\includegraphics[width=0.7\textwidth]{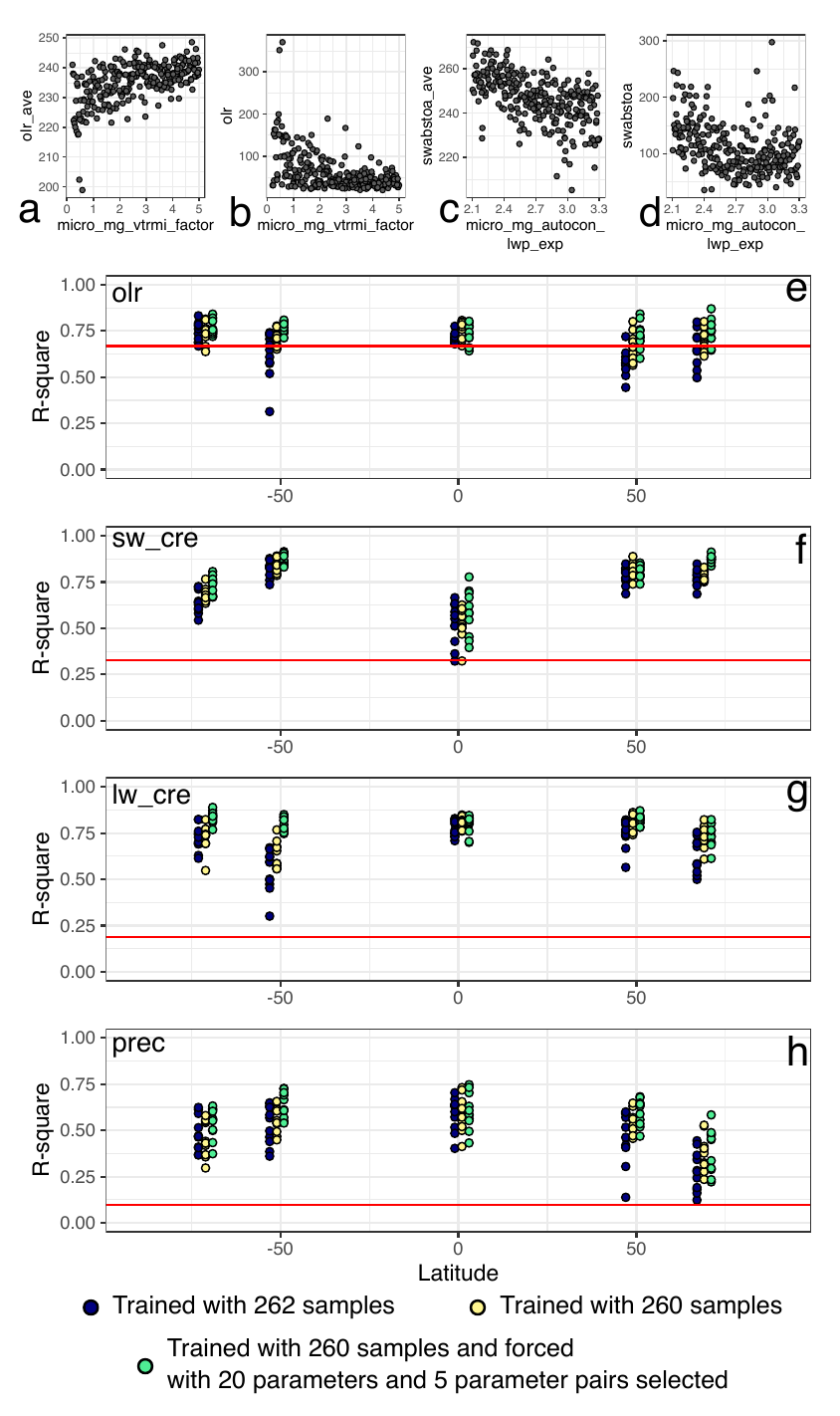}
    \caption{a-d: target variables ``olr$\_$ave'' (model output), ``olr'' (model score), ``swabstoa$\_$ave'' (model output), and ``swabstoa'' (model score) plotted against their most important parameters, respectively, from the CAM6 PPE Default Test; e-g: the R-squares of selected variables at different latitudes for the CAM6 PPE. The red horizontal lines denote the R-squares of the corresponding variables in the Default Test. The blue, yellow, and green points correspond to different ways to train the emulator. See text for more details. Their vertical spread denotes the variability from random sampling. Each cluster of the blue, yellow, and green points corresponds to the variable at the same latitude. Their x-axes are shifted slightly to allow for better visualization. }
    \label{f_cam_rmse_by_lat}
\end{figure}

Some of our target variables are model scores as opposed to raw model output. The calculation of the model scores compresses data at different latitude bands into one value (Eq. 1 in \citeA{elsaesser2024}). This is done so that the systematic uncertainty from satellite observations can be easily taken into account in a global bulk penalty score, with minimization of a bulk score being a more tractable effort for parameter estimation efforts  (see \citeA{elsaesser2024} for more details). But, this comes at a cost: a weakened relationship between the parameters and target variables. This scenario appears to be less of an issue in the ModelE3 PPE, but is present in CAM6. Moreover, as mentioned earlier, the model score is a function of the squared difference between the model output and observation, which further implies a loss of information about the model output. This can be  seen in Fig. \ref{f_cam_rmse_by_lat}a-d in which the  patterns between the model output and parameters become less evident when output is converted to scores.

There are a few other considerations possibly important for emulator performance.  First, if a few outliers exist in the PPE, their inclusion in the training or validation dataset will affect the emulator performance. This is especially true for PPEs with fewer samples. For example, Fig. \ref{f_cam_rmse_by_lat}a shows the presence of two ensemble members with extreme values for the variable \textit{olr} in the CAM6 PPE.

Additionaly, as suggested from our analysis of the ModelE3 PPE, a collection of individually less important parameters, when considered together, could have a cumulative impact on emulator performance. The small PPE size might make their effects harder to detect in the case of the CAM6 PPE. If we force \textit{sage} to select additional parameters, there is a chance that the emulator performance can be improved.

\subsection{Experiments yielding improvement in emulator performance}
We investigate to what extent addressing the factors discussed above affect the emulator performance. We select four variables and emulate the corresponding model outputs at five different latitudes (-71$^\circ$, -51$^\circ$, 1$^\circ$, 49$^\circ$, 69$^\circ$). The variables are those exhibiting different emulator performances in the Default Test (R-squares ranges from 0.09-0.66 therein). Three different ways of training are employed: (1) standard training as described in the method section; (2) exclusion of the two outliers in Fig. \ref{f_cam_rmse_by_lat}a, and subsequent training and testing of the emulator; and (3) exclusion of the two outliers, and forcing \textit{sage} to select 20 individual parameters and five parameter pairs during training. The latter is employed because, otherwise, it tends to select fewer parameters (less than 15 in all cases) and parameter groups (two to three in all cases), which might omit parameters that contribute information to the emulator. We do not force selection of parameter groups of three, since in the first two cases of training they were rarely selected in the parameter sequences.

Similar to what is done before, eleven tests (with the two outliers taken out in the second and third ways of training) are performed for each way of training the emulator and for each variable at each latitude.

The R-squares of the test results are plotted in Fig. \ref{f_cam_rmse_by_lat}e-h by the latitude. The performance of \textit{sage} for the variable \textit{olr} at different latitudes and with different ways of training the emulator is similar to its performance in the Default Test. The overall emulator performance improves noticeably for the other three variables at the selected latitudes compared to the Default Test.

The three different ways of training the emulator lead to slightly different performance. The R-squares from using the first way tends to have greater R-square variability (e.g., \textit{olr} at the latitude of -51$^\circ$ and \textit{prec}, precipitation, at 49$^\circ$). The second way of training in most cases leads to a slightly improved method performance compared to the first. Forcing more parameters and parameter groups (the third way of training) in the parameter sequences slightly improves the emulator performance for some variables, e.g., \textit{olr} and \textit{lw$\_$cre} (top of the atmosphere long wave cloud radiative forcing) at some latitudes.

Fig. \ref{f_cam_rmse_by_lat}f (\textit{sw$\_$cre}, top of the atmosphere short wave cloud radiative forcing) shows that the emulator performance varies at different latitudes. We also note that the parameters and parameter groups order in the parameter sequences varies by the latitude for the same target variables, suggesting their spatially-varied dependency on different simulated processes and the parameters that govern these processes. These suggest the potential of \textit{sage} in drawing more interpretations from climate model PPEs when applied to zonal or local model output.

With the lessons learned above, we calculate the global averages of 12 variables (Table \ref{t_table1}) in CAM6 PPE using \textit{sage} and the NN, trained on 210 ensemble members. The two outliers are excluded, and 11 experiments are performed. The resultant R-squares are presented in Fig. \ref{f_cam_rsq}c, where improvement in performance using both emulator methods is noted. The NN in this case has a slightly better performance.

\section{Discussion}
As an emulator, our new method is shown to perform comparably to a fully connected NN when applied to the two PPEs. Howver, the new method also quantifies the importance of parameters and parameter groups during the training and validation. Additional insights on emulator design are drawn from comparing the performance of the two methods.

\subsection{Differences from previous works}
The technical differences between \textit{sage} and other similar methods (e.g., additive GP and polynomial-based methods) are listed in the method section. \textit{sage} is unique in that analysis of a PPE and output of useful diagnostics is done simultaneously with emulator training. Other methods (e.g., \citeA{williamson2013history,sexton2019finding}) have the capability of quantifying parameter interaction, but such information is, in many cases, not automatically analyzed (e.g., the finding that it is not always optimal to emulate all target variables all at once). It is also common to examine the importance of  parameters and parameter interaction using sensitivity analysis (e.g., Shapley value or the Sobol's indices), which requires generating new data points based on the trained emulator (e.g., \citeA{li2019reducing}). The generated data contain uncertainty from the imperfect emulation which might affect the interpretation when the emulator performance is poor. \textit{sage} visualizes and outputs the variability explained by parameters and parameter interaction, providing more insights on the PPE without the need to implement additional analysis. \textit{sage} also outputs how much variability that cannot be explained, reducing the chance of over-interpretation.

Parameter interaction has been more explicitly studied in \citeA{bellprat2012objective}, but their emulator is prescribed to be of quadratic form, which is less flexible than \textit{sage} and does not support characterizing the interaction between more than two parameters. Their sampling scheme is chosen to better fit to the emulator format (and hence not from LHS), and the parameter space is of much lower dimensions (five). Their work also serves for objective calibration of model parameters which is not the focus of the present work. The above factors distinguish their work from the current work.

We consider the proposition of Eq. \ref{eq_measure_pair}, the metric or criterion to select the important parameter groups (parameter pairs in the particular equation), an innovation in this work. It considers the \textit{additional} benefit of emulating based on a group of parameters jointly relative to emulating them independently. If needed, we can use Eq. \ref{eq_measure_pair} to directly characterize parameter interaction without the need to emulate the effects of individual parameters (i.e., skip steps a-d in Fig. \ref{f_flowchart}).

\subsubsection{Comparison across two methods and two PPEs}
The application of two emulator methods (i.e., \textit{sage} and NN) to two PPEs and the separate analyses on them inform us of complementary advantages of the two methods and a wider perspective of how model parameters affect the model output. The comparison, analysis and findings from this work are considered valuable contributions to the study of climate model PPEs as they extend beyond one or two particular methods and have not been pointed out previously. 

Our work shows that \textit{sage} could have performance comparable to a NN but without the necessity of estimating the GP hyperparameters. We can therefore suggest that the relationship between the parameters and target variables learned by both methods is similar, and therefore characterized by low-frequency variability and interaction of only up to a few parameters. We do not need a sophisticated emulator or statistical model to capture such a relationship.

\subsection{Parameter importance conditioned on the parameter ranges}
The performance of an emulator is dependent on the nature of the PPE being analyzed. Even though emulating the model scores is likely to degrade the emulator performance, \textit{sage} and NN both exhibit improved performance for the ModelE3 PPE compared to the CAM6 PPE. Besides more obvious factors such as the number of ensemble members and differences in the climate models, the different performance is also related to how the parameter ranges are specified. Whether a parameter is important or not is thus conditioned on the ranges of itself and other parameters. This argument is somewhat intuitive, but its implication might worth attention: two PPEs generated from the same climate model but with different specified parameter ranges could lead to different or even conflicting interpretations.

\section{Summary and conclusions}
In this work, we present a new method, referred to as \textit{sage}, that implements analysis and emulation of climate model PPEs simultaneously. This method draws on a simplified version of additive Gaussian Processes, and caters to a key feature of climate model PPEs: a very limited number of ensemble members sampling a high-dimensional parameter space. \textit{sage} identifies and ranks the parameters and parameter groups that affect the value of a target variable, and models their impacts on the target variable as additive terms.

\textit{sage} is applied to two climate model PPEs (ModelE3 and CAM6), and it performs comparably to a NN.  Regarding ModelE3, improved performance is tied to an increase in the number of ensemble members, with further enhancement after accounting for the relationship between target variables during training. For CAM6, \textit{sage} slightly outperforms or has comparable performance (target variable dependent) to a NN. In this work, \textit{sage} is applied to emulate the global averages or averaged zonal skill scores of climate model outputs, although it can be used to emulate zonal or local climatologies.

Upon analyses of the emulator applied to ModelE3 and CAM6, we learn that the following points are critical for emulator design and emulator-facilitated analyses of PPEs:
\begin{enumerate}
    \item The relationship between the target variables and parameters captured by an emulator is dominated by the effects of individual parameters and parameter pair interactions. The relationship is also best characterized by low-frequency variation (as opposed to noise-like high frequency variation) of the target variable with respect to the parameters. This may be a universal feature of sparse climate model PPEs.

    \item Neither emulating the variables all at once nor one-at-a-time is (always) the perfect emulator design. Having separate emulators focusing on separate groups of target variables might lead to improved overall performance compared with one emulator simulating all variables at once. The presence of a stand-alone relationship between one or a few target variables and some parameters could be difficult to capture by a NN and potentially other emulation methods if all variables are emulated at once. Incorporating a relationship between target variables into the emulator, if a clear one exists, could further improve emulator performance. 

    \item A group of individually less important parameters could have a non-negligible and cumulative impact on the overall emulator performance. Thus, we need to be \textit{very} careful in determining whether a parameter is important or not. 

    \item Emulator performance may not always increase (linearly) with the number of training ensemble members (e.g., \textit{olr$\_ave$} in Fig. \ref{f_size_matter}a), and this is target variable-specific. It is therefore beneficial to generate ensemble members progressively (e.g., 100 at a time), followed by quantification of how performance changes with the presence of more training data, and then a decision on whether to generate more ensemble members. 

    \item How we define the target variables (i.e., for a given variable, do we formulate as a model score or raw output?) to be emulated may affect emulator performance. There are trade-offs to be made between directly emulating PPE member performance skill versus raw output. 
\end{enumerate}

Although we found less sensitivity to GP hyperparameter settings, which is an appealing feature of a machine learning emulator, \textit{sage} may be further improved by adopting a more rigorous way to optimize hyperparameters.  Work is currently underway to provide emulator uncertainty in a robust and realistic way, and such an enhancement will be described and provided in our publicly available emulator. Enhancements move us toward our overall goal of providing an emulator that enables the following: 1. diagnostics that provide insight on the relationship between parameters and outputs without underlying assumptions of linearity; 2. an emulator that can be reliably used in model automated parameter calibration efforts, and 3. an emulator that can provide reliable sensitivities of outputs to parameters across a variety of regimes and alternate parameter settings.

\appendix
\pagebreak

\section{Simple example of how GP hyperparameters affect emulation}
We use a simple example here to illustrate how the range and nugget-to-variance ratio affect the emulated mean of a GP. We randomly sample 100 points from the function $y = \sin(\pi x) + \sin(2\pi x) + 0.5*\sin(16\pi x)$ with x in the range of $(0, 1)$. We assume the first two terms as the signal and $0.5*\sin(16\pi x)$ as high-frequency noise in the data. The predicted GP means from using four pre-specified ranges and two nugget to variance ratios are presented as colored lines in Fig. A1b-d. Curves in b and c all have the same nugget to variance ratios of 1.0 and 0.5, respectively. Comparing these curves suggests that the value of nugget to variance ratio has limited impact on the predicted mean (i.e., the difference in b and c is small). Range of 1.0 leads to poorly-fitted results (light blue lines in b and c), ranges of 0.4 and 0.1 approximate the signal (i.e., $y = \sin(\pi x) + \sin(2\pi x)$) relatively well, and are not greatly different from each other (Fig. A1d) regardless of the different specified hyperparameters. With the specified range of 0.02 which is too small, the GP captures both the signal and high-frequency noise. 
\begin{figure}[H]
    \includegraphics[width=0.7\textwidth]{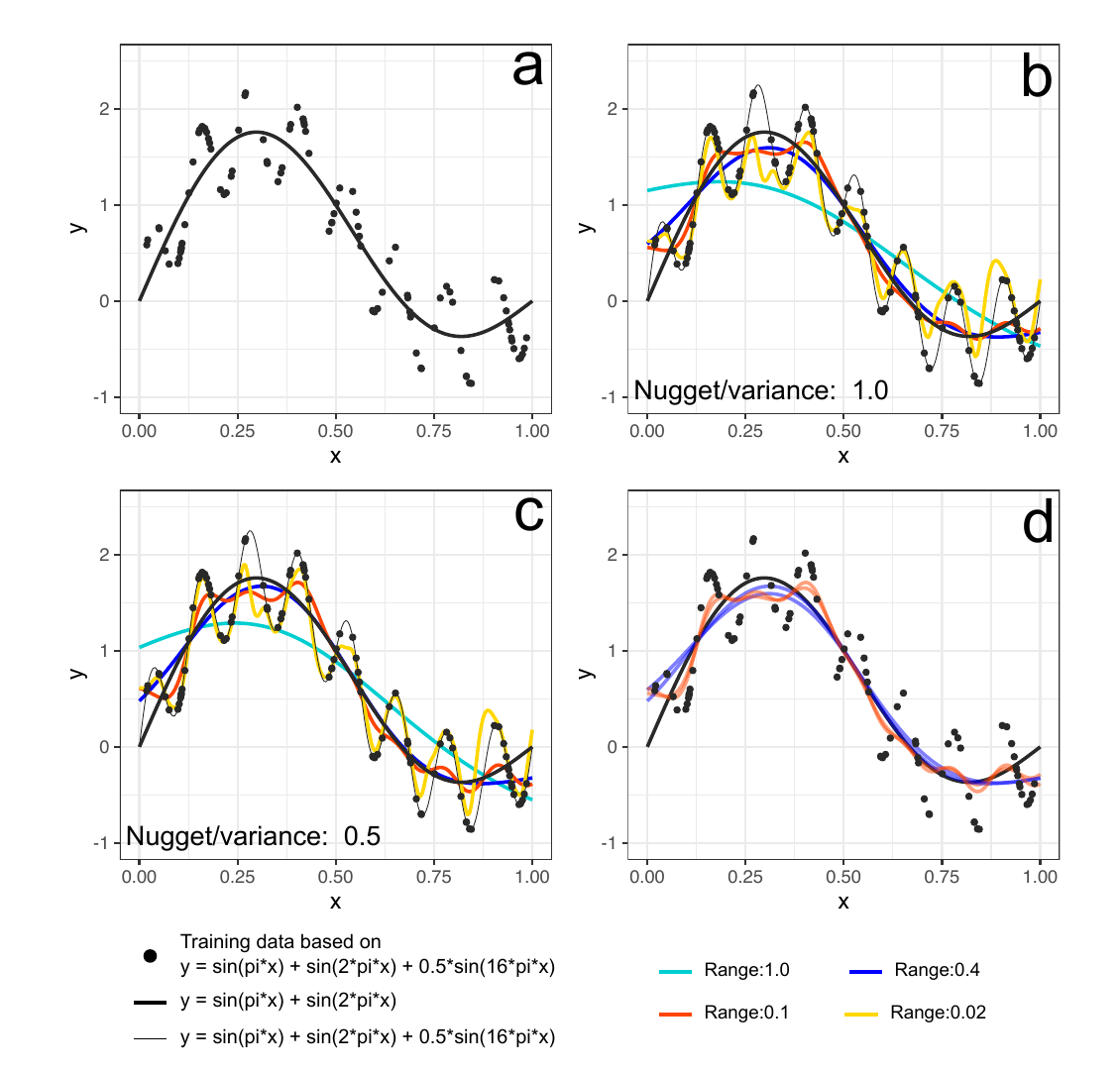}
    \caption{Black points sampled from $y = \sin(\pi x) + \sin(2\pi x) + 0.5*\sin(16\pi x)$. The thick black line in a-d marks the assumed signal in $y = \sin(\pi x) + \sin(2\pi x)$ and $y = \sin(\pi x) + \sin(2\pi x) + 0.5*\sin(16\pi x)$, the signal plus noise, marked as thin line in b-c. GPs with different ranges and nugget to variance ratios are fitted to the points with the emulated mean presented as colored lines in b and c. The difference between b and c is the specified nugget to variance ratios (1.0 and 0.5). Within each figure, the curves are fitted with different specified ranges. In d, four curves (ranges of 0.4 and 0.1; nugget to variance ratio of 1.0 and 0.5) that fit relatively well to the signal are presented, which are also similar to each other. }
    \label{f_appendixA_example1}
\end{figure}

\pagebreak
\section{Performance of a NN and the more conventional, non-additive GP in the Default Tests}
R-squares from using the NN and the more conventional, non-additive GP (only GP mean taken into account) in the Default Tests of the two PPEs are presented here. It is shown that the NN slightly outperforms the more traditional, non-additive GP.  
\begin{figure}[H]
    \includegraphics[width=\textwidth]{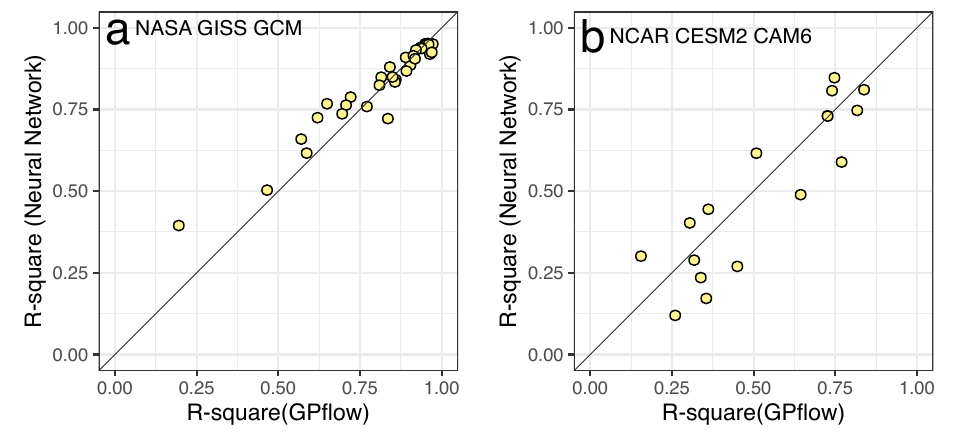}
    \caption{R-square comparison between NN and the more conventional, non-additive GP using GPflow in the Default Tests of the two PPEs.}
    \label{f_appendixB_giss_gp_nn}
\end{figure}

\pagebreak
\section{Explained variability in two random sampling tests for the ModelE3 PPE}
Explained variability classified and summed based on the sixth and seventh random sampling tests of the ModelE3 PPE using \textit{sage}. The classification criteria are identical to Fig. \ref{f_both_rmse_summary}a. Appendix C is presented to ensure the robustness of our interpretation of Fig. \ref{f_both_rmse_summary}a.

\begin{figure}[H]
    \includegraphics[width=\textwidth]{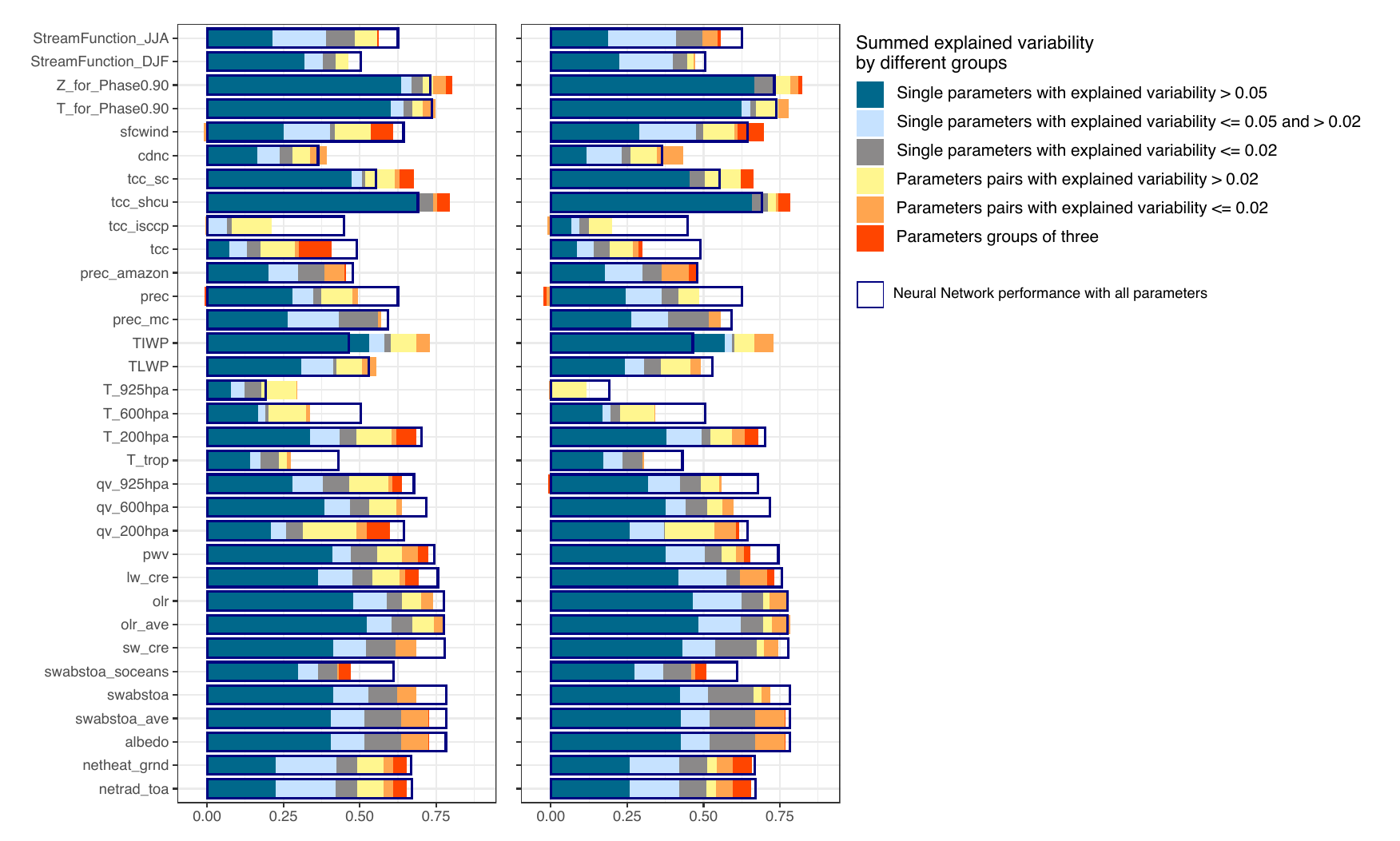}
    \caption{Explained variability classified and summed based on the sixth and seventh random sampling tests of the ModelE3 PPE using \textit{sage}. The classification criteria are identical to Fig. \ref{f_both_rmse_summary}a.}
    \label{f_appendix_D}
\end{figure}

\pagebreak

\section{Explained variability from using \textit{sage} with the consideration on the relationship between target variables}
Appendix D is presented to show the variability explained by individual parameters and parameter groups from using \textit{sage} with the consideration on the relationship between target variables.

\begin{figure}[H]
    \includegraphics[width=\textwidth]{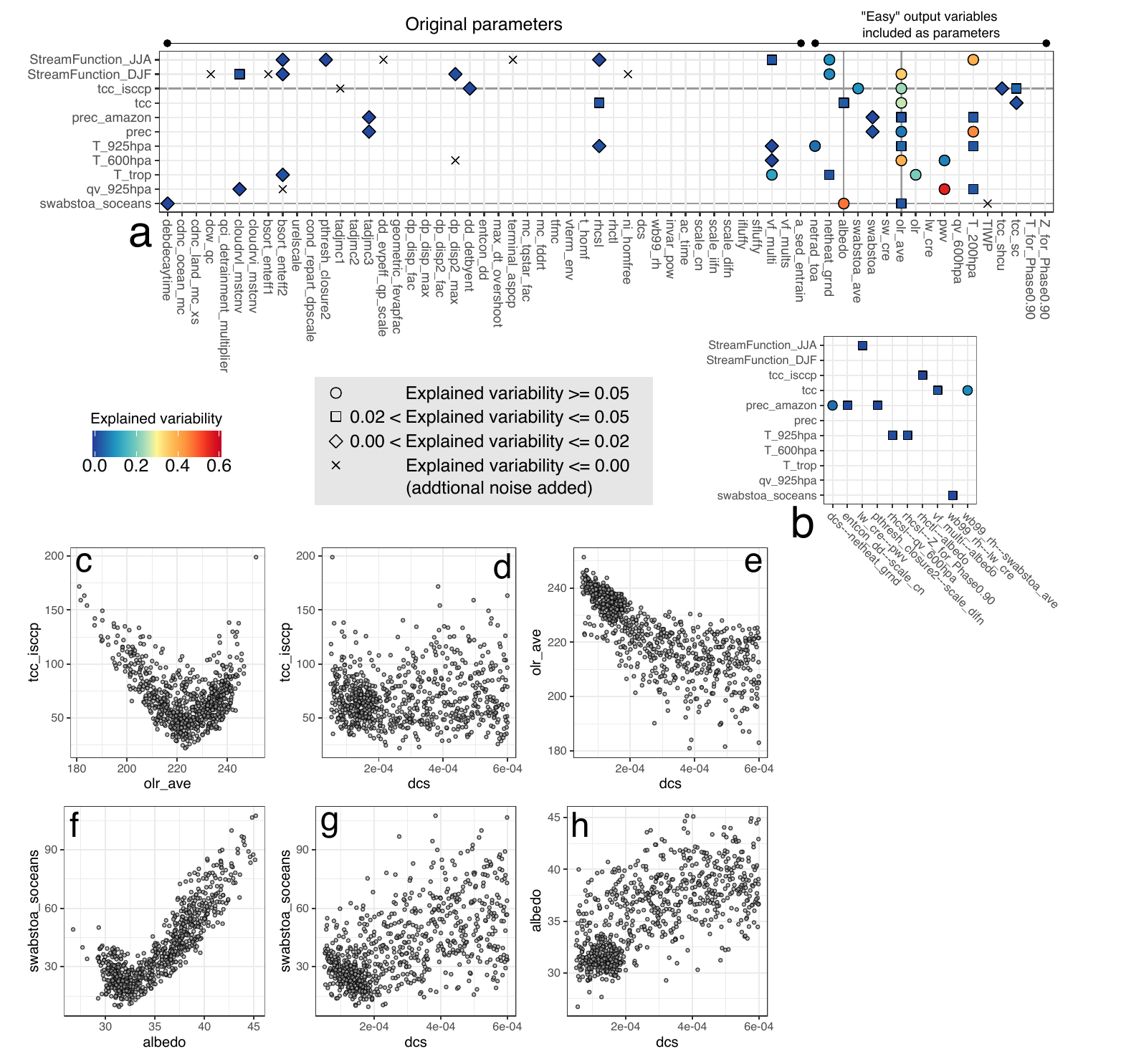}
    \caption{Variability explained (a and b) for the emulator design that considers the relationship between target variables for the ModelE3 PPE. Target variables that NN is better at emulating are considered here (green points in Fig. \ref{f_giss_rsq}a). Training and testing datasets in the Default Test are used here. a and b denote the contribution from individual parameters and parameter groups, respectively; c-h: bi-plots of selected parameters and target variables. }
    \label{f_giss_boost_rmse_summary}
\end{figure}

\pagebreak

\section{Example showing the relationship between model parameter, model output, and model score}
A simple example showing how the relationship between target variables could help improve the performance of \textit{sage}. We assume $x$ is a model parameter, $y1$, $y2$, $y3$ are the direct model outputs, and $y2\_score$ and $y3\_score$ are model scores of $y2$ and $y3$. Note that the relationship between $y3$ and $x$ is straightforward (i.e., a near straight line rather than the V-shape shown in the relationship between $y3\_score$ and x), and hence not shown here for simplicity.  

\begin{figure}[H]
    \includegraphics[width=\textwidth]{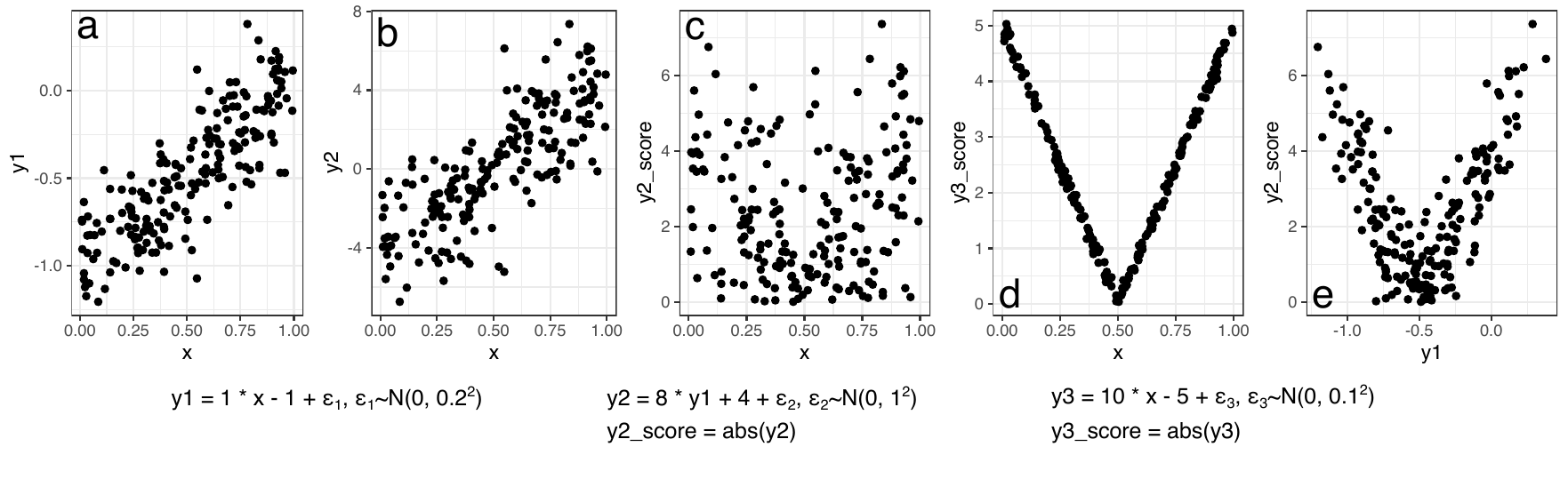}
    \caption{y1 (a) is a model output that depends on parameter x, and y2 (b) is another output that depends directly on y1 and thus implicitly on x. y2$\_$score is the model score of y2. The relationship between x and y2$\_$score is not significant (c). Whereas the relationship between y1 and y2$\_$score is more noticeable (e), which would lead to better emulator performance if such a relationship is considered. y3$\_$score is another model score (d), which shows that the model score is not always difficult to emulate, as a result of much lower noise level. The difference in c, d and e comes from the scale of the Gaussian noise. }
    \label{f_appendixE}
\end{figure}

%%%%%%%%%%%%%%%%%%%%%%%%%%%%%%%%%%%%%%%%%%%%%%%

\section*{Open Research Statement}
The datasets used in this work can be found in the supporting documents of this work. The codes and datasets used in this work are also made available on github (\url{https://github.com/yiqioyang/ppesage}).

\acknowledgments
We thank the editor and reviewers for their time, effort, and patience handling and commenting on our manuscript. We thank K. Loftus for providing valuable insights on our manuscript. We acknowledge funding from NSF through the Learning the Earth with Artificial intelligence and Physics (LEAP) NSF Science and Technology Center (STC) (Award \#2019625).  We acknowledge additional support from NASA-MAP (including grants \#80NSSC21K1498 and \#80NSSC17K1499 ).

\bibliography{main}
\end{document}